\documentclass{IEEEtran}

\usepackage{amsmath}
\usepackage{graphicx}
\usepackage{cite}
\usepackage{lscape}
\usepackage[colorlinks,allcolors=black]{hyperref}
\usepackage{caption}
\usepackage{subcaption}
\usepackage{tcolorbox}
\usepackage{booktabs} 
\usepackage{array}    
\usepackage{enumitem} 
\usepackage{svg}

\newtcolorbox[auto counter,list inside=mybox]{mybox}[3][]{%
    title={Box~\thetcbcounter:~#2},
    #1,
    halign title=center,
    sharp corners,
    fonttitle=\bfseries\sffamily\large,coltitle=black,titlerule=0pt,
    colbacktitle=white,
    colback=white,
    label={#3},
}

\title{What Really is `Molecule' in Molecular Communications? The Quest for Physics of Particle-based Information Carriers}
\author{Hanlin Xiao, \IEEEmembership{Student Member,~IEEE},
        Kamela Dokaj, \IEEEmembership{Student Member,~IEEE},
        and Ozgur B. Akan,~\IEEEmembership{Fellow,~IEEE}
       \thanks{H. Xiao and O. B. Akan are with the Internet of Everything (IoE) Group, Electrical Engineering Division, Department of Engineering, University of Cambridge, Cambridge, CB3 0FA, UK (email: hx289@cam.ac.uk, oba21@cam.ac.uk).

        O. B. Akan and K. Dokaj are also with the Center for neXt-generation Communications (CXC), Department of Electrical and Electronics Engineering, Koç University, Istanbul, 34450, Turkey.

        This work was supported in part by AXA Research Fund (AXA Chair for Internet of Everything at Koç University).

} 
}

\begin{document}
\maketitle

\begin{abstract}
Molecular communication, as implied by its name, uses molecules as information carriers for communication between objects. It has an advantage over traditional electromagnetic-wave-based communication in that molecule-based systems could be biocompatible, operable in challenging environments, and energetically undemanding. Consequently, they are envisioned to have a broad range of applications, such as in the Internet of Bio-nano Things, targeted drug delivery, and agricultural monitoring. Despite the rapid development of the field, with an increasing number of theoretical models and experimental testbeds established by researchers, a fundamental aspect of the field has often been sidelined, namely, the nature of the molecule in molecular communication.

The potential information molecules could exhibit a wide range of properties, making them require drastically different treatments when being modeled and experimented upon. Therefore, in this paper, we delve into the intricacies of commonly used information molecules, examining their fundamental physical characteristics, associated communication systems, and potential applications in a more realistic manner, focusing on the influence of their own properties. Through this comprehensive survey, we aim to offer a novel yet essential perspective on molecular communication, thereby bridging the current gap between theoretical research and real-world applications.


\end{abstract}

\section{Introduction}

\IEEEPARstart{M}{olecular} communication is a novel communication paradigm that relies on the exchange of information molecules for information transfer, in contrast to electromagnetic waves in traditional communication. It stands as a complementary approach to traditional electromagnetic communication, especially in environments like underwater or within the human body, where standard methods might struggle\cite{farsad2016comprehensive, akan2016fundamentals}. 


Specifically, molecular communication demonstrates the potential for nanoscale information transfer within the human body. It is widely regarded as the most promising approach for nanonetworks\cite{akyildiz2008nanonetworks} and the Internet of Bio-Nano things\cite{akyildiz2015internet, kuscu2021internet}. In these applications, nanomachines\footnote{Robots in microscale can also be used for the applications while employing molecular communication\cite{kong2018micro}. The network of micro/nanorobots is  referred to as nanonetwork in this survey, while nanomachines are historically used mostly.} form networks within the human body, collaborating to execute integrative functions such as targeted drug delivery\cite{chude2017molecular}, health monitoring\cite{nakano2020applications,atakan2012body, khan2020nanosensor}, and interfacing with external devices\cite{nakano2014externally,kisseleff2016magnetic}. Currently, various microrobots have been constructed showing the potential for targeted drug delivery\cite{xin2021environmentally, li2023overview}. Additionally, by applying the paradigm on natural biological systems, such as neuronal communication\cite{lotter2020synaptic, ramezani2018information, ramezani2017rate} and calcium signaling \cite{bicen2016linear}, we have the potential to gain a deeper understanding of the natural mechanisms that could profoundly improve the diagnosis and treatment of diseases such as spinal cord injuries\cite{akan2021information, civas2020rate}, neurological disease\cite{barros2018multi}, and olfactory diseases\cite{Dilara2023odor}

The field of molecular communication has experienced significant growth. Over the last decade, numerous reviews explored various facets of this domain. For instance, \cite{farsad2016comprehensive} offers a comprehensive overview of recent advancements, while \cite{jamali2019channel} centers on the modeling methods of Molecular Communication via Diffusion (MCvD). Additionally, \cite{bi2021survey} presents a hierarchy of research stages and examines current progress within that framework, \cite{chude2017molecular} presents a comprehensive survey on the application of molecular communication and molecular network paradigms to targeted drug delivery, \cite{akan2016fundamentals} provided a broad overview of the fundamentals of the connection between molecular information and communication science and \cite{soldner2020survey} the biological building blocks of synthetic molecular communication systems are reviewed. 

\begin{figure}[htp]
    \centering
    \includegraphics[width=0.5\textwidth]{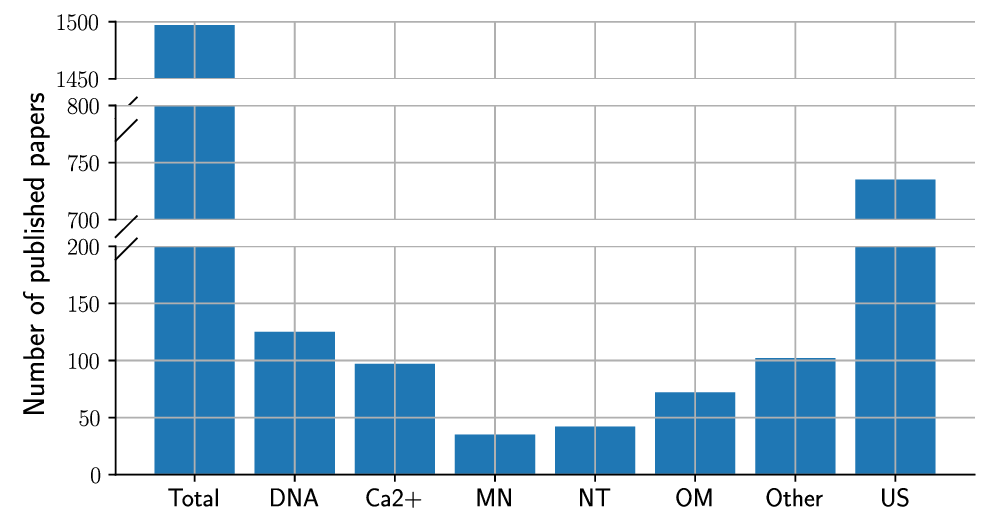}
    \caption{Distribution of published papers by information molecule type in Molecular Communication research. Data sourced from IEEE Xplore and ACM databases reveals that papers with unspecified information molecules take up the largest proportion. Abbreviations used: DNA, Ca$^{2+}$ ions, Magnetic Nanoparticles (MN), Neurotransmitters (NT), Odor Molecules (OM), and Unspecified (US).}
    \label{fig: papernum}
\end{figure}

While the field of molecular communication has seen rapid growth, a crucial aspect that is frequently sidelined is the information molecules themselves which can be shown by the large proportion of the unspecified papers in Figure \ref{fig: papernum}. Unlike electromagnetic waves in traditional communication, which are thoroughly studied in physics and governed accurately by the Maxwell equations, the information carriers in molecular communication — \textit{the molecules} — have an incredibly diverse range of properties such as subtypes, magnetic susceptibility, diffusivity, biocompatibility, etc. As the molecular communication channel directly depends on the physical particles moving from one point to another, these properties can greatly influence the setup of the communication channel, the channel performance, and the application scenarios of the molecular communication systems. However, in much of the existing literature, the information molecules are taken for granted with a lack of in-depth discussion about their properties. The distinct traits of different molecules are not fully captured by the channel models or leveraged to benefit the communication channel.
In this paper, we, for the first time, approach molecular communication by focusing on the information molecule as a primary factor. Our aim is to underscore the critical role these molecules play in actual molecular communication systems and to provide information that could spur further research. In this way, we hope to ensure that interested researchers have adequate information about the molecules to build more realistic theoretical models and experimental testbeds, thereby facilitating the transition of research to practical applications.

In order to guide our extensive survey, we have first identified the classes of molecules used in the literature, including DNAs, magnetic nanoparticles, calcium ions, neurotransmitters, odor molecules, and other less-explored molecules as presented in Figure \ref{fig: papernum}, and a straightforward visualization of these molecules can be seen in Figure \ref{fig: size comparison}. The paper is then structured into sections on these molecules. In each section, one class of information molecule is introduced. To provide a direct comparison between the different information molecules, the structure of each section is kept consistent along the features summarized below:
\begin{enumerate}
    \item 
    \textit{Physical Characteristics ---}
    First, we introduce the fundamental properties of information molecules, encompassing aspects including types, mass, charge, magnetic susceptibility, dimensions and diffusivity, and biocompatibility. These factors are pivotal for molecular communication channels. Specifically, dimensions and diffusivity affect the propagation of the information molecule in the medium. A plot for the direct comparison of the sizes of the information molecules is shown in Figure \ref{fig: size comparison}. Generally, the diffusivity $D$ of a spherical molecule through water can be obtained by the Stokes-Einstein relation, i.e.,
    \begin{equation}
        D=\frac{k_\text{B}T}{6\pi\eta r},
        \label{eqn: Einstein}
    \end{equation}
    where $k_\text{B}$ is the Boltzmann constant, $T$ is the temperature, $\eta$ is the viscosity of the liquid and $r$ is the radius of the particle\cite{nelson2008biological}. Note that equation (\ref{eqn: Einstein}) assumes a spherical molecule, so the shape of the molecules could add another layer of complexity to the diffusivity of the molecules.
    Moreover, the electric and magnetic fields could be used to control the propagation\cite{cho2022electrophoretic, bartunik2022planar}, and biocompatibility would determine if the information molecule could be used for in-body applications. Additionally, we delve into specific properties of molecules, such as the hybridization of DNA molecules. These details are essential for understanding the intricacies of molecular communication systems.

    \begin{figure*}
        \centering
        \includegraphics[width = 7in]{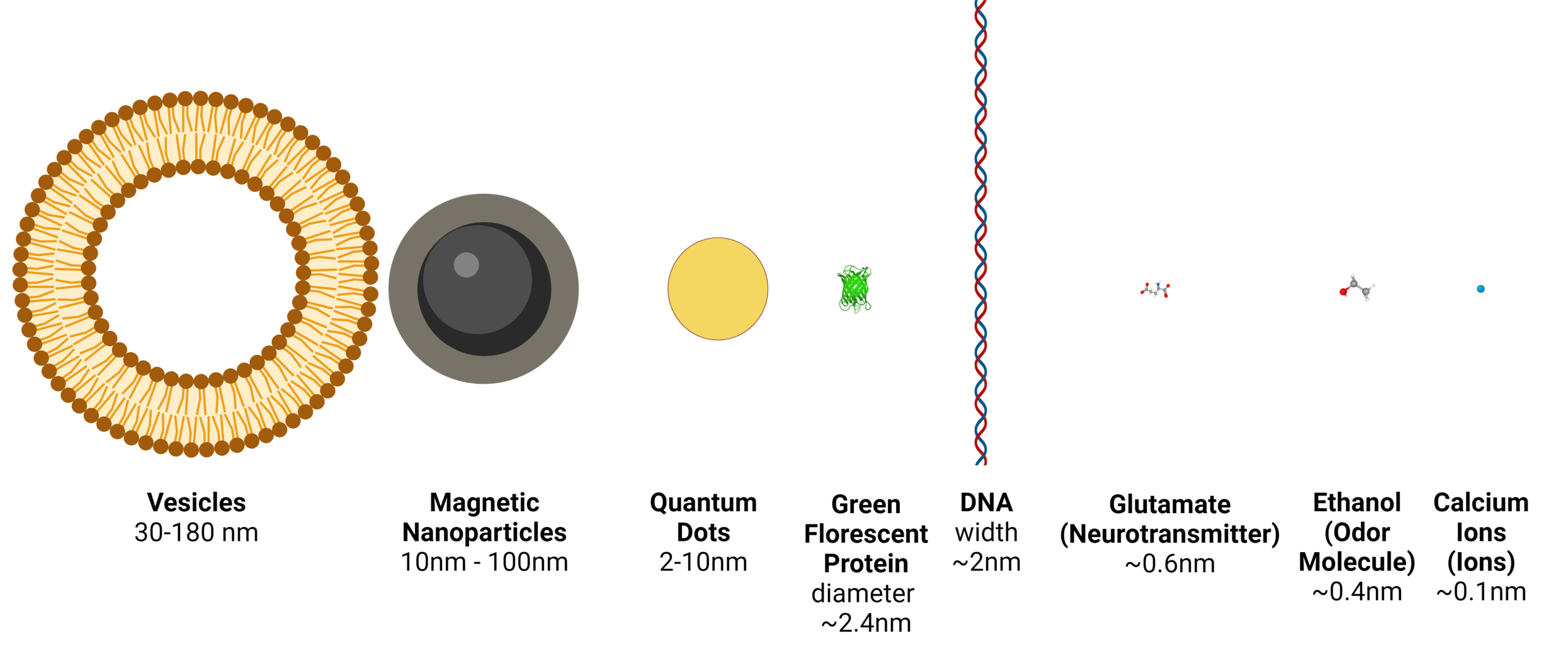}
        \caption{The figure illustrates the varying dimensions and structures of information molecules referenced in the paper, highlighting how their diverse sizes and forms influence propagation in the communication channel\cite{figure}. The size of the green fluorescent protein is cited from\cite{hink2000structural} and the sizes of the other molecules are discussed in the paper. Aerosols, notably larger at several micrometers, are excluded from this comparison. The 3D structures of glutamate and ethanol are cited from \cite{pubchem_2019b,pubchem_2019}.
        }
        \label{fig: size comparison}
    \end{figure*}
    
    \item 
    \textit{Communication Channels and Techniques ---}
    In this subsection, we explore the components of the communication channel, including the transmitter, propagation channel, and receiver, alongside the modulation methods employed for a specific information molecule class.

    The transmitters and receivers in molecular communication systems can be broadly classified into two categories: biological and artificial. Within the biological category, both naturally occurring organisms and biologically engineered (synthetic) organisms are considered. Natural organisms are examined to apply information-communication theoretic (ICT) metrics to biological systems. This approach offers a new perspective and enhances our understanding of these systems. Conversely, biologically synthetic components, along with artificial components, are explored in the context of experimental testbeds and for future implementations of molecular communication systems.
    
    Notably, in \cite{soldner2020survey}, an exhaustive proposal, and summary of possible designs for transmitters and receivers, particularly focusing on cations, neurotransmitters, and phosphopeptides, are provided predominantly from a synthetic biology standpoint. However, in this survey, we primarily concentrate on the components investigated in the existing literature. The reason is that these models are studied with more in-depth quantitative and qualitative details.

    Concerning the propagation channel, there exist three primary mechanisms for the propagation of information molecules: \textit{bacteria-based}, \textit{molecular-motor-based}, and \textit{diffusion-based} propagation which are elaborated later. Specifically, for the diffusion-based channel, three distinct scenarios arise in different physical environments. Firstly, the \textit{free diffusion} channel occurs in unobstructed, static space, where molecules move from regions of high to low concentration due to Brownian motion. The change in concentration distribution follows Fick's law of diffusion. Secondly, in the presence of flows such as wind or water flow, molecules are carried directly by the flow, augmenting free diffusion and forming a \textit{diffusion-with-advection} channel. Thirdly, if additional factors such as enzyme reactions are present in the channel, degrading the information molecules, an added layer of complexity emerges, resulting in a \textit{diffusion-with-reaction} channel. Mathematical modeling of these channels involves incorporating different boundary conditions and channel properties, and for detailed mathematical formulations, readers are referred to \cite{jamali2019channel}, which offers a comprehensive tutorial on diffusion channels.
    
    Regarding modulation methods in molecular communication, two predominant techniques are widely employed\cite{pehlivanoglu2017modulation}. The first is \textit{concentration-based modulation}, where the communication system operates within fixed time slots. Within each slot, the transmitter emits a pulse of information particles at a specific concentration, encoding information through particle concentration. For instance, \textit{On-off keying (OOK)} represents a simple version of concentration-based modulation, where the presence of a particle pulse (concentration $>$ 0) signifies logic high 1, while its absence (concentration = 0) indicates logic low 0. OOK is favored in many molecular channel analyses due to its reliability and simplicity. The second modulation method is \textit{type-based modulation}, where information is encoded based on particle types. For example, in a system with two types of information molecules, transmitting type one represents logic high 1, while type two signifies logic low 0. Particle categorization can be based on properties like length, size, or chemical composition. Apart from these techniques, molecular communication systems employ time-based modulation, space-based modulation, molecular mixture shifted modulation \cite{jamali2023olfaction}, and hybrid techniques. For a thorough understanding of these modulation methods in molecular communication systems, readers are encouraged to refer to \cite{kuran2020survey}, which provides a comprehensive survey of various modulation techniques. Additionally, \cite{shrivastava2021transmission} offers valuable insights into both static and mobile nanonetworks and their respective modulation schemes, particularly in the context of the Internet of Bio-Nano Things.

    
    \item 
    \textit{Communication Performance ---}
 In this section, we present and analyze the achievable data rates estimated for natural biological systems using the established model, as well as the data rates achieved by experimental testbeds. Moreover, we discuss other common communication metrics such as the bit error rate (BER), delay, and power efficiency of the communication system. Finally, we propose the primary sources of noise for the communication channel discussed in the section.

    \item 
    \textit{Communication Applications ---}
   In this subsection, we delve into the potential application scenarios for information molecules, first categorizing the communication range into four distinct scales: nanoscale (nm to $\mu$m), microscale ($\mu$m to mm), mesoscale (mm to m), and macro-scale (m to km) as summarized in Table \ref{tab: info moles}. Correspondingly, these applications typically fall into two broad categories. The first involves enhancing our understanding of biological mechanisms, which aids in the diagnosis and treatment of diseases\cite{Dilara2023odor,akan2016fundamentals,egan2023toward, akan2021information}. The second category involves the use of information molecules in artificial networks composed of nanomachines, i.e., nanonetworks\cite{akyildiz2008nanonetworks, akyildiz2015internet}. For example, nanonetworks are envisioned for in-body real-time monitoring and targeted drug delivery \cite{nakano2020applications}. Furthermore, with a molecular communication interface connected to external devices, the concept of the Internet of Bio-Nano Things (IoBNT) \cite{akyildiz2015internet, kuscu2021internet} can be realized. This concept aligns with the broader paradigm of the Internet of Everything \cite{dinc2019internet}, where data from nanonetworks can be integrated into a larger network encompassing everything, leading to expansive applications. In \cite{chude2017molecular}, the author has further classified the in-body scenarios molecular communication can work in, including the \textit{cardiovascular}, \textit{extracellular space/cell surface}, \textit{intracellular}, \textit{whole-body}, and \textit{nervous signaling channels}. The suitability of the information molecules in these channels will be discussed as well.

\end{enumerate}

    The structure of the remainder of the paper is outlined as follows. In Sections II to VI, a comprehensive discussion of widely studied information molecules is presented in parallel. This includes an exploration of DNA, magnetic nanoparticles, calcium ions, neurotransmitters, and odor molecules. Section VII provides a general overview of other information molecules that have been investigated within the molecular communication paradigm. Section VIII proposes a number of future research directions. Finally, the paper is concluded in Section VIII, summarizing the key findings and insights discussed in the preceding sections.
    
\begin{table*}[]
\centering
\caption{Summary of information molecules and their properties.}
\label{tab: info moles}
\resizebox{\textwidth}{!}{%
\begin{tabular}{c|c|c|c|c}
\hline
\textbf{Molecule}      & \textbf{Range of communication} & \textbf{Propagation channel} & \textbf{Modulation} & \textbf{Applications} \\ \hline
\noalign{\vskip 2mm} 
\hline
\textit{DNA} & Nano-micro scale & \begin{tabular}[c]{@{}c@{}}Diffusion, bacteria-based, \\ molecular-motor-based\end{tabular} & Concentration, type-based & \begin{tabular}[c]{@{}c@{}}Micro/nanomachine communication, \\ Data storage and transmission over long range\end{tabular} \\ \hline
\textit{Magnetic nanoparticles} & Nano-meso scale & \begin{tabular}[c]{@{}c@{}}Diffusion with advection \\ (liquid flow/magnetic force)\end{tabular} & \begin{tabular}[c]{@{}c@{}}Concentration, type, \\ spatial-based\end{tabular} & \begin{tabular}[c]{@{}c@{}}Micro/nanomachine communication, \\ in body/external interface\end{tabular} \\ \hline
\textit{Calcium ions} & Nano-micro scale & Diffusion & Concentration-based & Calcium signalling, Micro/nanomachine communication \\ \hline
\textit{Neurotransmitters} & Nano-micro scale & Diffusion with interaction & Concentration, type-based & Synaptic communication, Micro/nanomachine communication \\ \hline
\textit{Odor molecules} & Macroscale & Diffusion with advection(airflow) & Concentration, type-based & Long range communication \\ \hline
\end{tabular}%
}
\end{table*}

    \section{Deoxyribonucleic acids (DNAs)}
\begin{figure}
            \centering
            \includegraphics[width=3.5in]{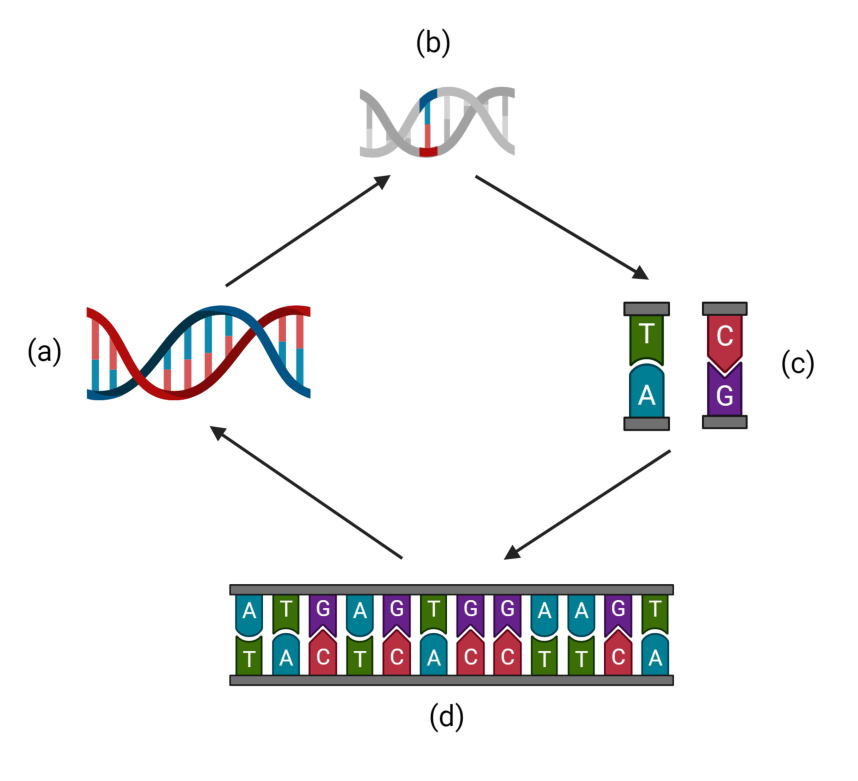}
            \caption{Depiction of the DNA molecule's architecture, illustrating its double-stranded, helical framework composed of complementary base pairs: (a) displays the double helix structure, (b) zooms in on a single base pair within the molecule, (c) identifies the specific base pairs, adenine (A) with thymine (T) and cytosine (C) with guanine (G), (d) shows a sequence of these base pairs, which form the genetic code within the DNA molecule.}
            \label{fig: DNA}
        \end{figure}

    Deoxyribonucleic acid (DNA) serves as life's foundational information storage system. It stores the genetic information that guides the development, functioning, growth, and reproduction of all known living organisms and many viruses. As shown in Figure \ref{fig: DNA}, each DNA molecule is a long, double-helical structure composed of two DNA strands made up of nucleotide units, each carrying one of four nitrogenous bases, i.e., adenine (A), thymine (T), guanine (G), and cytosine (C). The nucleotides on the two stands are complementary, with A pairing with T, and G pairing with C. The order of these bases encodes a large amount of genetic information. Moreover, DNA is very stable and compact, making it an ideal medium for data storage. Naturally, it is widely considered a perfect choice for information carrying and storage. In several studies, DNA-based communication in synthetic cells has been accomplished and investigated, using fluorescent proteins as reception indicators \cite{joesaar2019dna,ortiz2012engineered}. Various communication channels have been constructed and evaluated, employing DNA as information carriers with different propagation mechanisms \cite{cobo2010bacteria,gregori2010new,bilgin2018dna,castorina2016modeling,sun2019channel}. Additionally, DNA storage and DNA computing have been considered in conjunction with DNA-based communication\cite{dong2020dna,bell2016digitally,  liu2021dna, organick2018random}. On the other hand, DNA has more potential uses in the Internet of Bio-Nano Things and biosensors due to its ability to bind objects through \textit{hybridization} and be engineered to attach to specific microorganisms, proteins, and exosomes. For example, DNA can be engineered to bind to by-products of cancer for cancer detection\cite{dong2020dna} or as a loading/unloading mechanism for molecular motor cargos\cite{hiyama2007autonomous}. In this section, these properties and uses of DNA will be discussed in the context of molecular communication in detail.

        \subsection{Physical characteristics}
        

\begin{enumerate}


    \item 
        \textit{Types ---}
        The DNA molecules are made up of a strand of nucleotide bases. There are four types of nucleotide bases (A, T, C, G), and each combination of these bases can represent a different DNA molecule, effectively resulting in an infinite variety of DNA types. 

        From another perspective, DNA could be distinguished by different lengths for type-based modulation\cite{bilgin2018dna}. In addition, the folding of the DNA could have different topologies, such as circular, linear, supercoiled\cite{mirkin2001dna}. The topologies of DNA could further affect their diffusion coefficient\cite{robertson2006diffusion} and alter the propagation channel property accordingly. Furthermore, structural patterns such as hairpin loops\cite{chen2017ionic} and motifs\cite{d2006dna} can be used to distinguish different DNA strands to construct different symbol alphabets for molecular communications.

        
        Moreover, the topological structure of DNA, which includes forms such as circular, linear, supercoiled, etc.\cite{mirkin2001dna} can influence the DNA's diffusion coefficient\cite{robertson2006diffusion}, consequently altering its propagation channel properties.  Additionally, fluorescent dyes have been used to label DNA\cite{proudnikov1996chemical, arthanari1998fluorescent}, which require a preprocessing of DNA molecules with a florescent dye solution. When using this method to label DNA, the biocompatibility of the florescent dyes, and their effects on the structure of the DNA need to be considered.
                
    \item 
        \textit{Mass ---}
        The mass of DNA molecules can be directly obtained from their length. Each base pair is about 600 Daltons, $1.08 \times 10^{-24}$ kg\cite{alberts2017molecular}. There are approximately 3 billion base pairs in each human cell, corresponding to $3.24 \times 10^{-15}$ kg. The small mass of DNA also makes them suitable for transmission at the microscale.
        
    \item 
        \textit{Charge and Magnetic susceptibility ---}
        DNA molecules are negatively charged due to the phosphate groups in their backbone. They are diamagnetic, which means they become weakly magnetized in the opposite direction of the applied magnetic field. This property is associated with electronic orbital motions within the DNA molecules\cite{yi2006emergent}. 
        Gel electrophoresis is a laboratory technique used to separate and analyze biomolecules, including DNA, RNA, and proteins. It uses an electric field to move these molecules based on their charge and size\cite{magdeldin2012gel}. Similarly, electric fields could potentially be used to assist DNA molecule diffusion in molecular communication systems, an area not yet explored in the existing literature.

    \item 
        \textit{Dimensions and Diffusivity ---}
        Regarding the dimensions of DNA molecules, there exists an additional layer of complexity due to their potential variance in length based on the number of base pairs they contain. A fully unraveled DNA strand from a single human cell can stretch up to 2 meters, containing approximately 3 billion base pairs. Yet, despite its considerable length, DNA is notably thin, with a width approximating 2 nm\cite{hardison2021working}. This slenderness allows for DNA to be intricately folded, enabling its storage within the limited volume of cells. The compactness combined with its remarkable information-bearing capability makes DNA an intrinsically prospective molecule for informational roles in nanomachines. Ongoing research delves into DNA nanotechnology's potential, examining its prospects in forming diverse nanostructures and devising molecular machines at the nanoscale\cite{park2022recent,dey2021dna}.
        
        Furthermore, the topology of DNA introduces additional complexity to its diffusivity\cite{robertson2006diffusion}. As elucidated in \cite{robertson2006diffusion}, it has been found that the diffusivity of DNA largely follows a power law, i.e., 
        \begin{equation}
        D = D_0 L^{-\nu_i},
        \end{equation}
        where $D$ denotes the diffusion coefficient, $D_0$ is a coefficient dependent on the medium, $L$ indicates the number of base pairs in the DNA, and $\nu_i$ is a coefficient influenced by the DNA's topology. This can range from linear, relaxed circular, to supercoiled forms. As a numerical example, in aqueous solutions, the diffusion coefficients of DNA fragments in water decrease from 53 $\times$ 10$^{-8}$ to 0.81 $\times$ 10$^{-8}$ cm$^2$/s for sizes of 21 to 6000 base pairs\cite{lukacs2000size}. Therefore, when modeling DNA-based molecular communication channels, the specific diffusivity of the DNA used needs to be considered carefully. Especially, when using length-shifting type-based modulation, the corresponding change of diffusivity will introduce extra noise in the communication channel.

    \item 
        \textit{Biocompatibility ---}
        While DNA is inherently biocompatible, its expression within cells can lead to potentially harmful effects and may trigger immune responses. Therefore, It is vital to engineer DNA to be inactive or non-expressive to mitigate unintended consequences\cite{gregori2010new}. 
        At the same time, since the transportation of the DNA can be achieved by not only diffusion but also bacteria-relay and molecular motor (this will be discussed later), the biocompatibility of the biological infrastructure needs to be considered as well. Escherichia coli (E. coli), being the typical bacterial carrier of the DNA, naturally resides in the human intestine. On the other hand, in molecular motor-based systems, key components like kinesin and microtubules, are naturally present in the human cells as well. Although harmful bacteria and genetically engineered kinesins and microtubules still exist, as long as every component is carefully selected and engineered, biocompatible DNA-based molecular communication channels can be achieved.
        

    \item 
        \textit{Hybridization ---}
        Single-stranded DNA (ssDNA) can transiently form during several cellular processes, including DNA replication, repair, and recombination. In the right salt and temperature conditions, ssDNA will spontaneously bind to a complementary ssDNA strand. This process is known as \textit{hybridization}. DNA-based molecular communication systems utilize this mechanism to facilitate connection and detection between ssDNA molecules\cite{dong2020dna,hiyama2007autonomous}.

        \end{enumerate}




        
        \subsection{Communication Channel and Techniques}
        Other than diffusion, DNA molecules can be actively transported by bacteria and molecular motors. Therefore, the communication models need to account for the different transporting mechanisms as well.

\begin{figure}
            \centering
            \includegraphics[width=3.5in]{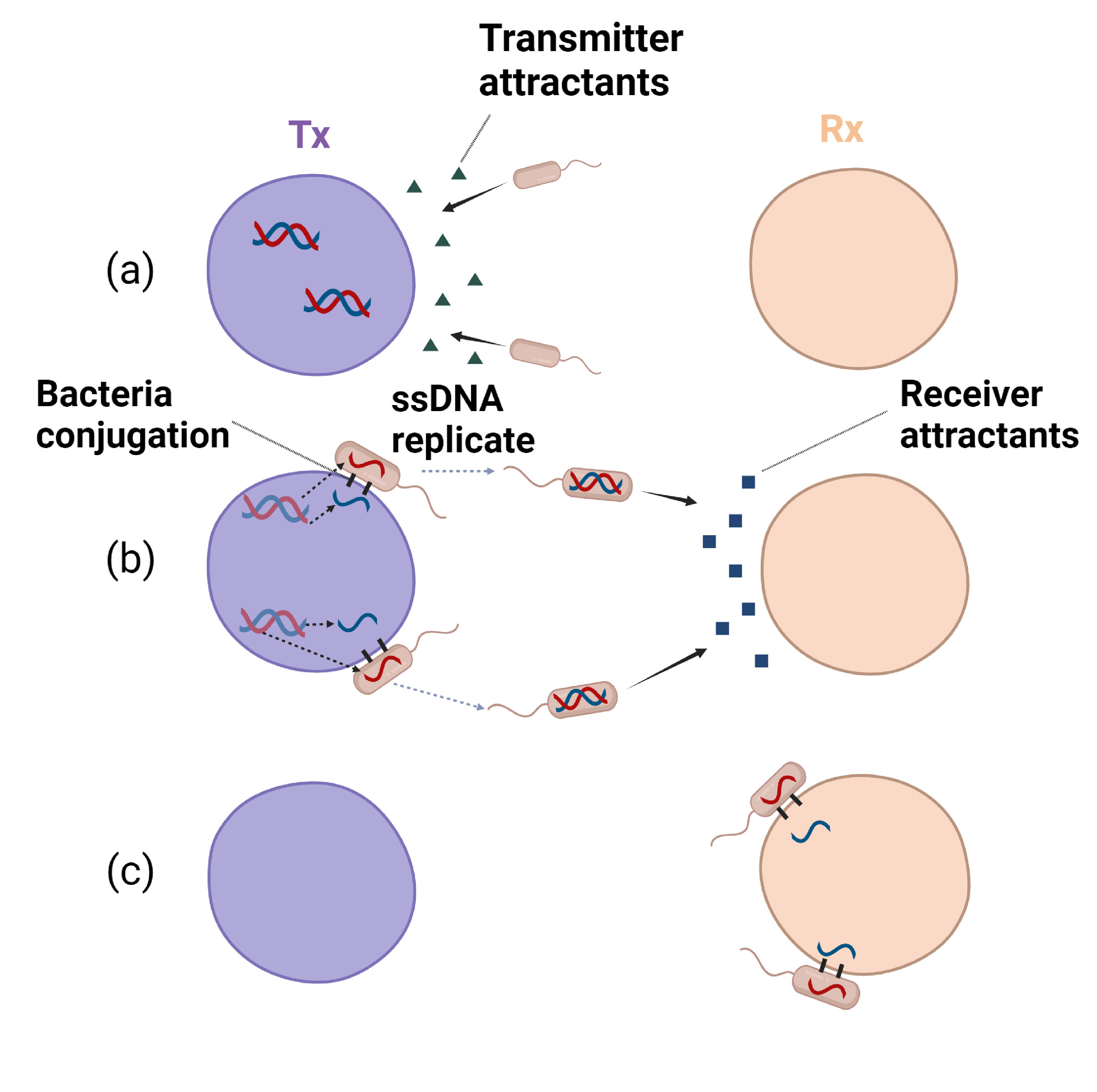}
            \caption{Schematic representation of intercellular communication via bacterial conjugation: (a) The transmitter node (Tx) emits specific attractants, guiding bacteria towards it. (b) Upon arrival, a single-stranded DNA (ssDNA) plasmid is transferred to the bacteria through conjugation, while concurrently, the receiver node (Rx) releases its own attractants to recruit the bacteria. As the bacteria migrate, the ssDNA plasmid within replicates, forming a double-stranded DNA (dsDNA) plasmid. (c) The bacteria, now carrying the complete dsDNA plasmid, reach the receiver node, where the plasmid is conveyed to the receiver through a subsequent conjugation event.}
            \label{fig: bacteria}
        \end{figure}

        In \textit{bacteria-based communication}, the transmitter is a DNA Processing Unit (DPU)\cite{cobo2010bacteria}. This DPU can synthesize a DNA plasmid—a small, circular, double-stranded DNA molecule with encoded information. These plasmids can be transferred to bacteria through a process known as \textit{bacterial conjugation}\cite{balasubramaniam2013multi}. During this process, a single-stranded DNA (ssDNA) detaches from the original plasmid and is transferred to the carrier bacteria with the aid of a pilus. To attract transporter bacteria, the DPU emits a transmitter attractant, prompting bacteria to move towards locations with higher concentrations of these attractants using their flagella—a behavior known as \textit{chemotaxis}. After acquiring the plasmid, the ssDNA within the bacteria regenerates its complementary strand, reconstituting a complete plasmid\cite{cobo2010bacteria}. The bacteria are then drawn to the receiver by an attractant emitted by it. Figure (\ref{fig: bacteria}) illustrates this entire procedure. This process can recur multiple times along neighboring nodes, serving as a relay system in bacterial nanonetworks\cite{balasubramaniam2013multi, qiu2017bacterial, gregori2010new}.

        In a \textit{molecular-motor-based system}, the communication system relies on microtubules moving along kinesin proteins. These proteins provide the driving force for the transport of cargo. The hybridization of single-stranded DNA (ssDNA) serves as the mechanism to autonomously load and unload this cargo\cite{hiyama2007autonomous,hiyama2007design,hiyama2010biomolecular}, as illustrated in Figure (\ref{fig: motor}). It is worth noting that, in addition to DNA, large liposome molecules containing other smaller objects, such as therapeutic and imaging agents\cite{vieira2016getting}, can also be transported by the molecular motor.
        

\begin{enumerate}

    \item 
        \textit{Transmitter ---}
        Currently, there are no well-established prototypes of nanomachines capable of autonomously synthesizing, storing, and transmitting DNA molecules. Nonetheless, examining certain aspects of existing models and technologies can provide insights into the construction of a DNA molecule transmitter. 


        The transmitter is composed of three key components: the information source, processing unit, and releasing unit, as depicted in Figure \ref{fig: transmitter}. A thorough review of molecular communication transmitter and receiver design can be found in \cite{kuscu2019transmitter}. Examining each component sequentially:
        \begin{enumerate}
            \item Information Source and Processing Unit: A DNA synthesizer can produce DNAs and encode information onto them using specific nucleotide base sequences. Over the past decades, DNA synthesis methods have transitioned from phosphoramidite chemistry to enzymatic synthesis\cite{dong2020dna, caruthers2013chemical}. In current experimental testbeds, researchers either synthesize the DNA themselves\cite{hiyama2007autonomous,hiyama2011micropatterning} or obtain it from suppliers like Integrated DNA Technologies\cite{joesaar2019dna,dong2020dna}. Furthermore, a cellular storage system model has been proposed to store and transmit DNA molecules for communication purposes\cite{shah2017molecular}. While energy harvesting techniques specific to DNA transmitters are not discussed in depth, general energy harvesting methods encompass mechanical, thermal, biocell, and electromagnetic means\cite{saraereh2020hybrid,zou2021recent}.
            
            \item Releasing Unit: While often depicted as a point-wise source \cite{bilgin2018dna}, more detailed distribution models tailored to specific transmitters should be developed for greater realism in future studies. Moreover, a number of specific physical designs of the releasing unit are discussed in \cite{soldner2020survey}, including pipettes, voltage-gated channels, etc.
        
        \end{enumerate}

        \begin{figure}
            \centering
            \includegraphics[width=3.5in]{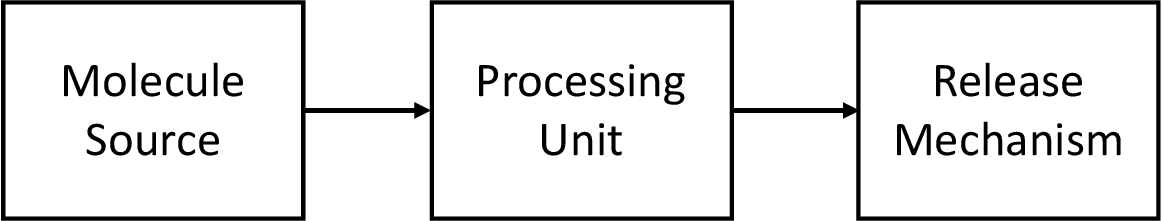}
            \caption{This schematic outlines the fundamental components of a transmitter system, illustrating the flow from the initial molecule source to the final release mechanism.}
            \label{fig: transmitter}
        \end{figure}

        


        \item 
        \textit{Propagation Channel ---}
        As highlighted earlier, DNA-based information molecule propagation is categorized into three primary mechanisms: diffusion-based, bacteria-based, and molecular-motor-based.
        
        


       \textit{Diffusion-based systems ---}
       In these systems, the propagation of DNA conforms to the diffusion equation, with advection potentially influencing the process based on particular system dynamics\cite{jamali2019channel,bilgin2018dna}. Factors impacting the diffusion of DNA include its size and topology, the viscosity of the surrounding liquid, any present flow within this environment, and the strength of external fields.
        
        \textit{Bacteria-based systems ---}
        The efficiency of DNA propagation in this system hinges on various bacterial properties, such as their quantity, movement speed, and sensitivity to attractants. In bacterial networks, bacteria approach each other either randomly via diffusion or purposefully through chemotaxis, opportunistically transferring plasmid information via bacterial conjugation. Comprehensive research on information propagation in bacterial relay networks is discussed in \cite{castorina2016modeling}.
        
        \textit{Molecular-motor-based systems ---}
        In these systems, ssDNA serves as a binder between the cargo and the microtubule-motor system. DNA hybridization triggers the binding and release of cargo. The cargo's movement relies on microtubule mobility over kinesin, but molecular-motor-based systems without ssDNA as binders also exist\cite{taira2006selective, diez2003stretching}. However, the ssDNA offers the benefits of selective, autonomous, and parallel cargo transportation\cite{hiyama2010biomolecular}.

        \begin{figure}
            \centering
            \includegraphics[width=3.5in]{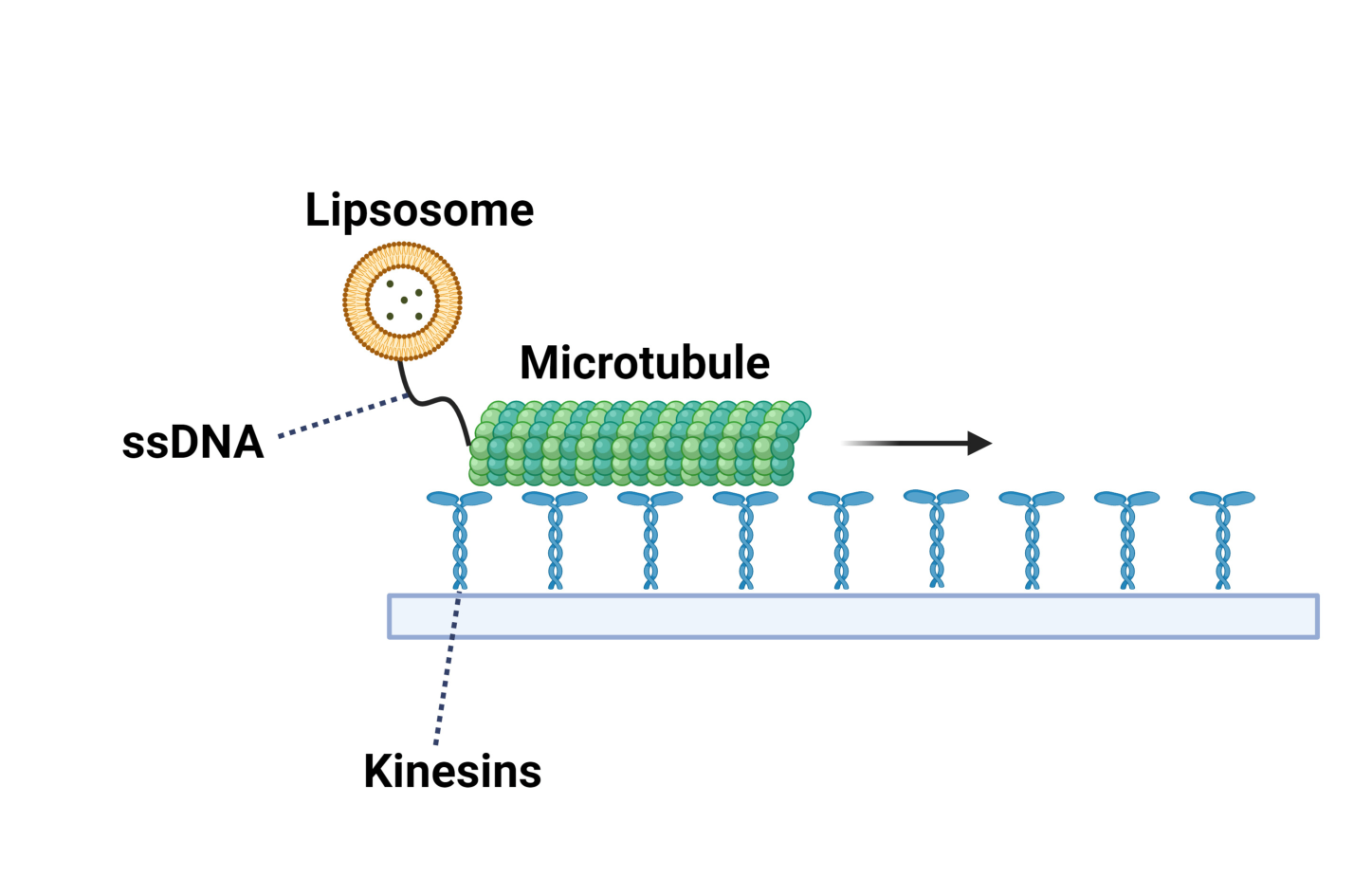}
            \caption{This schematic diagram illustrates an artificial molecular motor system. A microtubule, propelled by kinesins anchored to a substrate, transports a liposome cargo. The cargo is tethered to the microtubule through a single-stranded DNA (ssDNA) adaptor, facilitating the translation of the microtubule's linear motion into the directed movement of the cargo.}
            \label{fig: motor}
        \end{figure}

        \item 
        \textit{Receiver ---}
        While information is encoded within the bases of DNA or in the types of DNA, receivers must sequence the DNA or measure certain DNA properties. Currently, the most notable commercialized DNA sequencing technology measures changes in ionic current as individual DNA molecules translocate through a nanopore\cite{jain2016oxford}. Impressively, it can detect sequences up to 900 kilobases in length and is as compact as a USB disk\cite{shendure2017dna}. This technology has been employed in reading DNA-stored data\cite{dong2020dna,organick2018random} and is considered a detection method in molecular communication research\cite{bilgin2018dna}.

        However, in the present state of molecular communication experiments, seamlessly integrating a DNA sequencing unit into testbeds poses challenges. Instead, researchers use DNA fluorescence expression\cite{joesaar2019dna,tang2018gene} or the binding of fluorescent dye to DNA\cite{hiyama2007autonomous} to indicate DNA transfer, resorting to optical microscopy for detection in these instances.
        
        Another promising DNA molecule detector for molecular communication is the biological-Field-Effect-Transistor (bioFET)\cite{kuscu2016physical}. A conventional FET comprises a source, drain, channel, and gate that controls the current flow between the source and drain. When voltage is applied across the source and drain, current enters the source, passes through the channel, and exits via the drain (as shown in Figure \ref{fig: bioFET}). The channel, made of semiconductors, has electrical properties influenced by the connected gate. As such, any external stimulus to the gate can alter the channel's properties, thereby affecting the current/capacitance between the source and drain. In the case of a bioFET, the gate is replaced with a molecule recognition unit. Consequently, the presence of target molecules can be detected by the change in the current/capacitance of the bioFET, resulting from the recognition unit interacting with the molecule. Research in \cite{kuscu2021fabrication} presents a graphene-based bioFET that employs ssDNA as recognition units to detect ssDNA information molecules. The molecule's concentration encodes information for molecular communication. ssDNA is preferable due to its ability to bind with its complementary ssDNA through hybridization. Furthermore, they are versatile molecule recognition units since they can be designed to bind to peptides, proteins, carbohydrates, and other small molecules.

        \begin{figure}
            \centering
            \includegraphics[width = 3.5in]{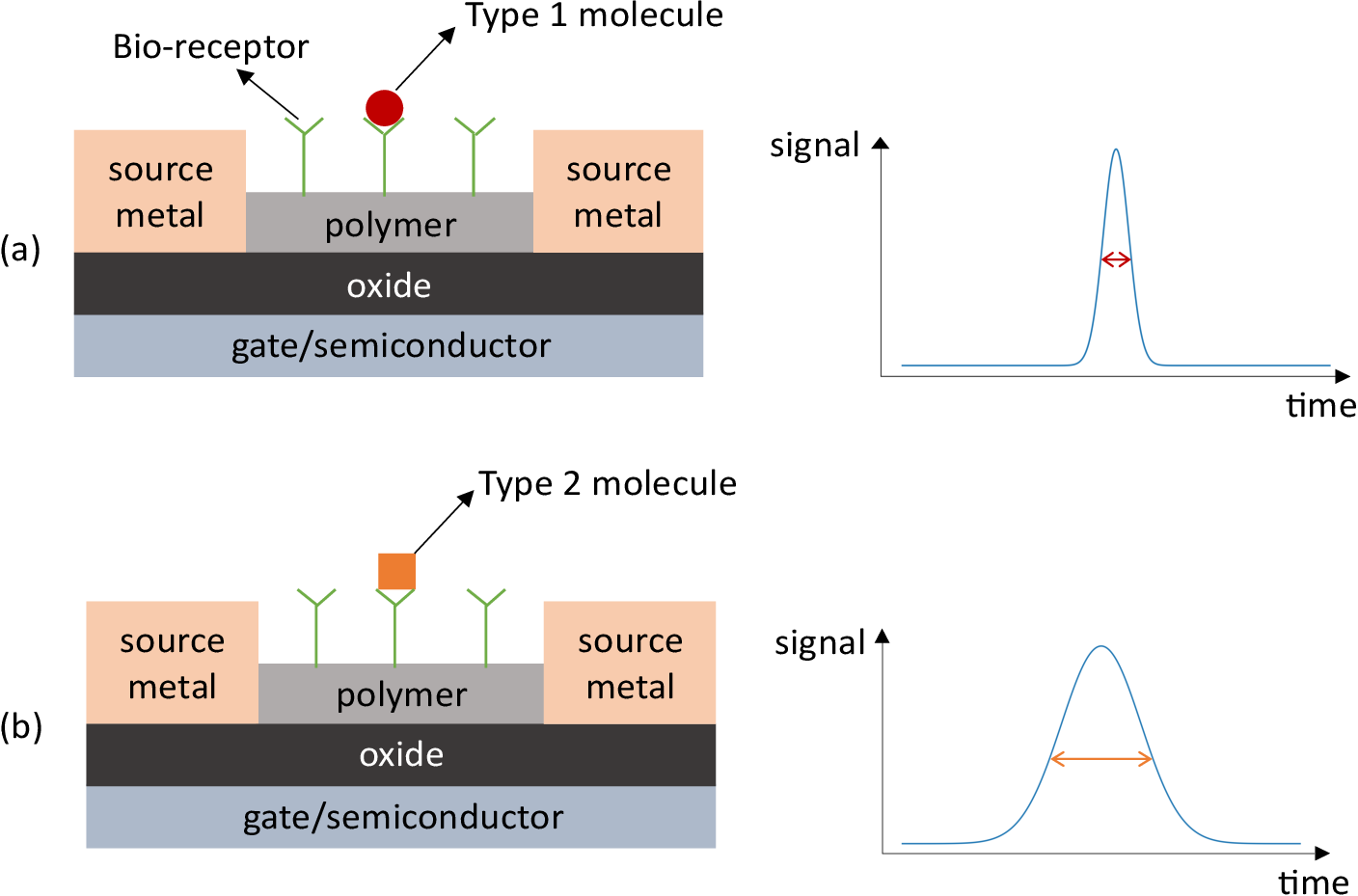}
            \caption{This diagram demonstrates the operational mechanism of a BioFET sensor. The BioFET comprises a gate region coated with bio-receptors (depicted as green Y-shaped structures) atop a field-effect transistor's surface. (a) shows the sensor upon capture of a molecule (red circle), with signal output depicted as a sharp peak. (b) illustrates the sensor upon the capture of another target molecule (orange square), which corresponds to a distinctive wide-peak wave signal on the BioFET's output.}
            \label{fig: bioFET}
        \end{figure}


        \item 
        \textit{Modulation ---}
        High-density information is ideally encoded by the sequence of nucleotide bases in DNA. Given that there are four possible bases (A, T, C, G), each base can represent 2 bits of information. For instance, if a binary encoding was used, A, T, C, and G could correspond to 00, 01, 10, and 11, respectively. This encoding means that a long DNA strand can store vast amounts of data within a relatively small volume.

        However, when envisioning the Internet of BioNano Things, it becomes apparent that integrating DNA sequencing technology into nanomachines is highly challenging. Given that nanomachines are anticipated to have basic functions, cooperating to achieve broader tasks, the dense data storage capacity of DNA molecules might not be particularly beneficial in such scenarios. Therefore, using nucleotide-based DNA modulation is more appropriate for macroscale, high-volume data transfer, even if it currently faces issues like slow read/write speeds\cite{dong2020dna}.
        
        In a nanonetwork, alternative modulations like concentration-based and type-based modulation become more practical. For example, in \cite{joesaar2019dna}, the florescent expression of DNA is employed to signal the arrival of DNA to the detector, and the signal could be picked up by microscopy. As such, OOK can be applied to this system, where the binary digits 0 and 1 are represented by the absence or presence of DNA within a given time frame. This modulation can be a feasible method for communication between nanomachines and external devices. Furthermore, in \cite{bilgin2018dna}, DNAs of varying lengths are utilized for type-based modulation. This is because the length of the DNA can be inferred from its translocation time through the nanopore. In addition to length, other DNA features, such as hairpin loops\cite{bell2016digitally} and sequence motifs\cite{chen2017ionic}, can also serve as markers for type-based modulation.

        Additionally, a number of encoding techniques have been developed to improve the performance of the DNA-based communication channels. In \cite{petrov2014forward}, the forward-reverse coding method is introduced to reduce the error rate in DNA communication by encoding information on both the forward and reverse strands of DNA molecules.

        Furthermore, \cite{walsh2010development} explores DNA-based molecular communication protocols inspired by current telecommunication standards. This study proposes a bi-layer structure that comprises both an \textit{encoding compartment }and a \textit{transmission and error Recovery compartment}.

        Additionally, in \cite{walsh2009hybrid}, protocol stacks for transmitting, routing, and receiving nodes are detailed. The paper investigates techniques for relaying information over intermediate addresses. An extrusive strand of ssDNA, capable of binding to the receiver, is utilized for address registration. A more comprehensive discussion about these protocols can be found in \cite{walsh2013protocols}.
        \end{enumerate}    

        \subsection{Communication performance}
        \begin{enumerate}

        \item
        \textit{Data Rate ---}
        In \cite{bilgin2018dna}, a diffusion-based DNA communication system is proposed with a potential data rate of 6 bit/s. Furthermore, \cite{hiyama2011micropatterning} conducts a channel capacity analysis for diffusion-based DNA communication, exploring the relationship between channel capacity, the number of base pairs in the DNA, communication distance, and time slot. Notably, many current experiments emphasize showcasing the capability of DNA for information transmission, often neglecting a comprehensive assessment from an information and communication theory (ICT) perspective. The field would greatly benefit from additional experimental testbeds that prioritize ICT performance analysis, potentially accelerating the development of this emerging paradigm.

        Also, in \cite{kuscu2021fabrication} the bioFET-based DNA communication system registered a data rate of 0.17 bit/s with a bit error rate(BER) of 5\%, and it is crucial to note that this system is a novel experimental testbed that can potentially be scaled down to the nanoscale.

        \item
        \textit{Noise Sources ---}
        Inter-symbol interference (ISI) remains a challenge for both diffusion-based DNA communication and bacterial networks. This is especially true for bacterial networks, where propagation involves opportunistic relaying, making the timing of arrival hard to predict precisely. A potential solution to this is to program the DNA to exterminate the bacteria\cite{cobo2010bacteria}.

        Additionally, incomplete conjugation processes among bacteria might introduce extra noise. One proposed solution is to integrate antibiotics into the nanomachines, which would target and eliminate bacteria lacking complete messages in their plasmids\cite{balasubramaniam2013multi}. Other prevalent noise sources include errors due to spontaneous DNA mutations\cite{sun2019channel}, and misidentifications that arise during DNA sequencing\cite{deamer2016three}.

        
        \end{enumerate}


        \subsection{Communication Applications}

        DNA-based molecular communication for the IoBNT can span various ranges, depending on the propagation mechanism.

        In the nanoscale and microscale, DNA-based molecular communication channels can be established through molecular motors\cite{alberts2017molecular} or diffusion\cite{bilgin2018dna}. A potential application scenario for this range involves communication between intercellular and intracellular nanomachines, while intracellular typically have a scale of 10 and 100 micrometers, the typical size of eukaryotic cells.
        
        Conversely, from microscale to mesoscale, relay bacteria networks might be more suitable\cite{gregori2010new, tsave2019anatomy}. The multi-hop technique in these bacterial networks makes communication over longer ranges viable\cite{balasubramaniam2013multi}. Although bacteria might have difficulties operating in the bloodstream due to immune system responses or potential infections, they are frequently active in other bodily environments. Bacteria are often found on mucosal surfaces like the gut or upper respiratory tract\cite{hou2022microbiota}. Thus, with careful engineering, bacteria can be envisioned to cooperatively form nanonetworks in these specific environments.

        Therefore, with the ability to work on multiple scales, DNA-based molecular communication would be suitable for working in in-body scenarios including cardiovascular, extracellular and extracellular spaces for the communication of micro/nano robots for task such as targeted drug delivery and health monitoring.


    
        Furthermore, DNA expressions, like fluorescence, offer a potential interface between nanomachines and external devices, serving as an \textit{outmessaging interface}\cite{nakano2014externally}.
        
        Lastly, DNA storage remains a vibrant area of research, drawing interest due to DNA's remarkable compactness and stability as a data storage medium\cite{dong2020dna}. This unique attribute suggests its potential for extended range and time of communication. By transporting encoded DNA physically via, e.g., vehicles, communication across vast distances and extended time durations could be accomplished.

        


    \section{Magnetic nanoparticles}

    With the first discussions of magnetism in medicine dating back to 1960\cite{freeman1960magnetism} and the subsequent advancement of nanotechnology, magnetic nanoparticles have attracted considerable interest within the medical community. Their potential applications span from enhancing medical imaging, specifically in Magnetic Resonance Imaging (MRI) where they serve as contrast agents to heighten image clarity\cite{shan2010superparamagnetic}, to innovative strategies in targeted drug delivery. By conjugating therapeutic agents, such as chemotherapy or radioisotopes, onto these nanoparticles and utilizing an external magnetic field, there is potential to guide the medicines precisely to the target sites. Such an approach could revolutionize treatments, offering increased efficacy coupled with reduced side effects\cite{wahajuddin2012superparamagnetic}. 
    
\begin{figure}
        \centering
        \includegraphics[width=3.5in]{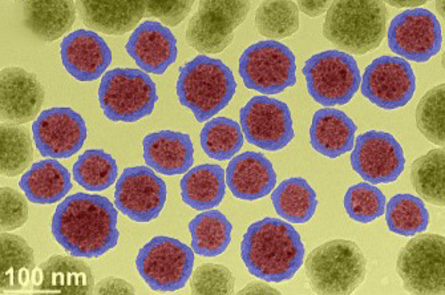}
        \caption{Microscopic image of Superparamagnetic Iron Oxide Nanoparticles (SPIONs), adapted from \cite{kopanja2016core}. Each particle consists of a magnetic core, typically composed of iron oxide, which is surrounded by a stabilizing coating. The core is visible as the red central region, while the coating may be represented by the blue outline. }
        \label{fig: SPION}
    \end{figure}

    Superparamagnetic Iron Oxide Nanoparticles (SPIONs) have attracted significant attention due to their unique superparamagnetism, which means they are magnetized only under the influence of an external magnetic field. In the absence of such a field, SPIONs do not remain magnetized, reducing the potential for unwanted agglomerations, and making them generally biocompatible. SPIONs are composed of an iron oxide core surrounded by a coating that can consist of polymers, small molecules, or proteins. The core imparts superparamagnetiam, while the shell serves multiple purposes. It enhances biocompatibility, provides stabilization\cite{thorek2006superparamagnetic}, and can be functionalized to offer specific features such as targeting ligands\cite{khaniabadi2020trastuzumab} and incorporating fluorescent dyes\cite{tiwari2020fluorescent}. A picture of their structure is provided in Figure \ref{fig: SPION}.
    Their small size, biocompatibility, and ability to be manipulated by external magnetic fields make them especially appealing for molecular communication research. Their potential applications span across intricate environments like the bloodstream and tumor sites. Their ease of production, combined with the aforementioned qualities, has invigorated the development of numerous theoretical models\cite{wicke2018molecular,kisseleff2016magnetic, wicke2019magnetic} and experimental testbeds\cite{bartunik2023development, wicke2021experimental} on SPION-based molecular communication. Meanwhile, the burgeoning interest in SPIONs has paved the way for research into device design tailored for their detection\cite{ahmed2019characterization,bartunik2022capacitive}.

        \subsection{Physical Characteristics}

        \begin{enumerate}
        
            \item
            \textit{Types ---}
            The core of SPIONs typically exists in two forms: $\gamma $Fe$_2$O$_3$ (maghemite) and Fe$_3$O$_4$ (magnetite). While they differ at the level of crystal structure, they present a practical challenge for differentiation due to their closely related crystal structures and magnetic properties. Techniques like Mössbauer spectroscopy\cite{winsett2019quantitative} and X-ray diffraction\cite{hiraga2021maghemite} have been employed to differentiate between them. However, the inherent complexities associated with these methods make them impractical for type-based modulation based solely on the SPION core types.
            
            In contrast, the coatings applied to SPIONs offer a more promising approach for type-based modulation. In the context of targeted drug delivery, the coatings of SPIONs can be meticulously tailored to bind to specific receptors\cite{wei2021superparamagnetic, kievit2011surface}. This tailored binding property holds the potential for type-based modulation. In molecular communication, however, much of the existing literature on SPION receivers primarily emphasizes their magnetization properties. This makes the differentiation between various types of SPIONs remain challenging.

            \item 
            \textit{Mass ---}
            The densities of Fe$_3$O$_4$ and $\gamma $Fe$_2$O$_3$ are approximately 5.2 g/cm$^3$\cite{materialsproject_mp-19306} and 4.9 g/cm$^3$\cite{webmineral-maghemite}. Therefore, with a known radius of the core, shell, and the density of the core, the mass of the SPION could be estimated. For example, a SPION core with a 10 nm radius would have a mass of 2 $\times$ 10$^{-20}$ kg. The mass of the magnetic nanoparticle will have multiple effects on the molecular communication channel. For example, we know that one major advantage of magnetic nanoparticles is that they can be actively controlled by an external magnetic field. But with a larger mass, the particle will be less affected by the magnetic force, which will further affect the delay of the communication, as well as the noise due to a weaker clearance of the channel.

            \item
            \textit{Charge and Magnetic Susceptibility ---}
            Even in the absence of an artificial coating, iron oxide nanoparticles inherently possess an oxide layer. The charge of this layer is intricately dependent on the pH of the surrounding medium\cite{sumner1963effect}. Notably, any coating applied to the nanoparticle can significantly alter the SPION's charge. Such variations in charge profoundly influence the adsorption characteristics of SPIONs with proteins\cite{sakulkhu2015significance}, i.e. the process of SPIONs adhere to the protein surface. Consequently, when designing SPIONs for molecular communication applications, it becomes imperative to meticulously evaluate their surface charge. However, much of the existing SPION-based molecular communication literature predominantly emphasizes in vitro experimental frameworks based on inorganic setups. As a result, the nuanced effects and potential applications of surface charge remain largely unexplored.  
            Regarding magnetic susceptibility, as discussed before, SPIONs exhibit superparamagnetism, meaning that they become magnetic in the presence of an external magnetic field. However, unlike ferromagnets, their magnetism vanishes once the external field is removed. This high magnetic susceptibility allows SPIONs to be used as contrast agents in MRI. Furthermore, it enables the manipulation of SPIONs within the body using an external magnetic field. Several factors can influence the magnetic susceptibility of SPIONs, such as their size, core material, and shell material\cite{rajan2020assessing}, as well as temperature\cite{strkaczek2019dynamics}. In a SPION-based molecular communication channel, the propagation of the SPIONs can be artificially controlled using a magnetic field which would be very useful in targeted drug delivery and in turbulent environments. Meanwhile, their presence can be detected by exploiting their magnetic susceptibility\cite{bartunik2022capacitive}. Therefore, to build a precise model of a practical SPION-based communication channel, the aforementioned factors all need to be carefully considered and incorporated into the channel models.



            \item 
            \textit{Dimensions and Diffusivity ---}
            Typically, SPIONs possess a hydrodynamic diameter (which accounts for both the core and the coat) of approximately 10nm-100nm. The ratio between the core and the shell may vary depending on the synthesis methods. The size of the core and shell can be tailored to achieve desired properties and performance in specific applications. For example, larger core SPIONs with effectively denser shells show weaker membrane adsorption and lower cell uptake\cite{gal2017interaction}. Also, during actual production processes, the SPIONs derived from coprecipitation often adhere to a log-normal distribution\cite{kiss1999new} in their size. The tiny, customizable size allows magnetic nanoparticles to be used for nano-scale communication. The diffusivity has been estimated in literature by Einstein relation (\ref{eqn: Einstein}) with a measured size of the particle (shell and core), temperature, and viscosity of the medium\cite{wicke2019magnetic, wicke2018molecular}. However, the shape of SPIONs can actually vary widely, and its determination is influenced by multiple factors such as the reaction conditions and the chemicals used in the synthesis process\cite{wahajuddin2012superparamagnetic}. Common shapes include spherical, nanoworms, rod-shaped, and magnetic beads\cite{mahmoudi2009cell}. For instance, in the coprecipitation method of SPION synthesis, a higher ratio of iron (indicating a low relative iron concentration) relative to polymers combined with elevated temperatures tends to yield smaller nanobeads. The shape and size of the nanoparticles influence their diffusion and biocompatibility. Generally, smaller, spherical SPIONs diffuse more easily due to their reduced size, but they might exhibit increased toxicity. This highlights an essential trade-off between diffusion efficiency and biocompatibility\cite{mahmoudi2009cell}. Therefore, the shape of the SPIONs remains a critical parameter to consider, especially in the context of molecular communication.



            \item
            \textit{Biocompatibility ---}
            SPIONs, comprising core materials such as maghemite and magnetite, offer inherent biocompatibility. As these nanoparticles degrade within the body, the resultant iron ions are not alien; they are integrated into the body's established iron metabolism pathway. This seamless integration is attributed to the pivotal role iron plays in various human physiological processes. However, while the core offers intrinsic biocompatibility, the overall compatibility of SPIONs in a biological environment often hinges on their surface coatings.
    
            The surface coating of SPIONs significantly influences their biocompatibility in biological systems. Coatings such as dextran and polyethylene glycol (PEG) not only enhance their stability but also reduce undesired interactions in the biological milieu\cite{zschiesche2022biocompatibility}. 
    
            Morphology also plays a role in SPIONs' biological interactions. Among various shapes, nanobeads have been observed to exhibit higher toxicity compared to their nanorod and nanosphere counterparts\cite{mahmoudi2009cell}. Additionally, there are concerns with ultra-small SPIONs. Their diminutive size allows for easy penetration through cell membranes, which might compromise intracellular structures and potentially induce toxicity\cite{wahajuddin2012superparamagnetic}. Therefore, when designing in-body SPION-based communication channels, the SPIONs have to be carefully engineered to avoid potential toxicity.
                
            \end{enumerate}

        
        \subsection{Communication Channel and Techniques}

        \begin{enumerate}
            
        \item 
        \textit{Transmitter ---}
        In the transmitter structure depicted in Fig \ref{fig: transmitter}, several components are pertinent to the use and management of magnetic nanoparticles, specifically SPIONs.

        The information source primarily functions as a storage unit for synthesized SPIONs. A variety of methods exists for the synthesis of SPIONs, ranging from co-precipitation and laser evaporation to thermal decomposition\cite{ali2021review}. Co-precipitation emerges as the predominant technique in molecular communication testbeds\cite{bartunik2023development,wicke2021experimental}. This method involves dissolving iron salts in water, followed by the introduction of an alkaline agent, which prompts the precipitation of iron oxide nanoparticles. Once formed, these nanoparticles are isolated, rinsed, and dried, preparing them for subsequent utilization.
        
       At the nanoscale, dedicated storage units for SPIONs have not been established. However, in macroscale experimental testbeds, mechanisms like micropumps and peristaltic pumps serve the dual roles of storage and emission for these nanoparticles\cite{bartunik2023development}.

        \item
        \textit{Propagation Channel ---} 
        SPIONs are primarily conceived to propagate within constrained diffusion channels, which might be influenced by both natural flow (as in the case of blood circulation) and externally applied magnetic fields. Therefore a diffusion-with-advection/reaction channel can be used as the model. Such channels are found in various regions of the human body. The inherent flow in these vessels could either facilitate or impede the propagation of SPIONs. Meanwhile, the introduction of an external magnetic field can be strategically used to enhance the performance of the communication channel by guiding the SPIONs. The intricacies of modeling such a diffusive molecular communication channel have been explored in depth in \cite{jamali2019channel}.

        \item
        \textit{Receiver ---}
        Different methods are being explored for the macroscale detection of Superparamagnetic Iron Oxide Nanoparticles (SPIONs). One such method harnesses an inductive coil that surrounds the propagation channel. As SPIONs traverse through the coil, their magnetic susceptibility induces a change in the coil's inductance\cite{ahmed2019characterization}. In another novel approach, a device has been crafted that uses a capacitor to discern variations in permittivity brought about by the presence of SPIONs\cite{bartunik2022capacitive}. Furthermore, by employing a detector array that integrates different detector types, both spatial and type-based modulations could be realized for SPION-facilitated molecular communication\cite{bartunik2022capacitive}. The miniaturization of these devices is crucial for practical applications. From another perspective, by tailoring the coating of SPIONs, they can be fashioned to bind to specific receptors. Such binding events can be subsequently ascertained using an array of bioanalytical techniques, converting the event into a discernible signal.

        \item
        \textit{Modulation ---}
        In the current literature, OOK is predominantly employed for SPION-based molecular communication, reflective of the nascent stage of this technology. However, as discussed above, there is potential to improve these preliminary methods. Type-based modulation, which uses distinct coatings as identifiers, and spatial modulation, utilizing a circular detector array, are emerging strategies that promise higher data transmission rates in SPION-based molecular communications.

         \end{enumerate}

        
        \subsection{Communication Performance}

        \begin{enumerate}

        \item 
        \textit{Data Rate ---}
        Achieving optimal data rates in magnetic-nanoparticle-based communication systems requires meticulous consideration of various influencing factors. These include the shape and size of the magnetic particles, the dynamics and geometry of the propagation channel, and potential attenuation and noise encountered during signal transmission. Furthermore, the efficiency with which SPIONs are transmitted and detected, as well as guidance from any external magnetic field, play crucial roles. To develop an accurate model of such communication channels, it is imperative to account for each of these factors and measure the corresponding parameters rigorously. The latest experimental testbeds, utilizing liquid diffusion-with-flow channels and operating without magnetic field guidance, have achieved a data rate up to 6.34 bit/second\cite{bartunik2023development}.

        \item 
        \textit{Noise Sources ---} 
        Intersymbol Interference (ISI) remains a substantial challenge in SPION-based molecular communication due to the overlap of subsequent symbol transmissions. In advection-dominant systems, the directed flow of the liquid medium or external magnetic field facilitates quicker removal of transmitted SPIONs, offering alleviation of ISI. In in-vivo environments, SPION-based communication encounters additional challenges, particularly due to macrophages which can ingest SPIONs, effectively causing attenuation in the communication channel. Using stabilizing coatings, like polyethylene glycol (PEG), can significantly decrease this cellular uptake, preserving signal strength. Beyond macrophage-related attenuation, turbulence in the propagation medium, variabilities in SPION synthesis, transmission methodologies, and detection processes all act as potential sources of noise. Turbulence, for instance, may unpredictably alter SPION movement, while inconsistencies in synthesis or transmission can affect signal uniformity. For SPION-based molecular communication to reach its full potential, continuous monitoring and mitigation of these noise sources are essential, encompassing strategies from periodic channel cleaning to sophisticated error correction techniques.

        \end{enumerate}





        \subsection{Communication Applications}

        Due to the small sizes and stability, SPIONs are able to work across nanoscale to mesoscale. At the nanoscale and microscale, the SPIONs have been investigated as drug carriers in targeted drug delivery\cite{mornet2006magnetic} and are also considered a promising information carrier for intrabody nanonetwork communication. In this way, in targeted drug delivery systems, SPIONs actually have a dual role. They can serve as drug carriers\cite{wahajuddin2012superparamagnetic}, as well as act as information particles for communication.\cite{chude2017molecular}. Furthermore, they are used in hyperthermia treatments for cancer patients, offering a promising therapeutic approach\cite{rajan2020assessing}. 
        On the mesoscale, SPIONs-base molecular experimental testbeds have already been implemented\cite{wicke2018molecular,kisseleff2016magnetic}. Therefore, SPIONs are able to work in many scenarios including cardiovascular, intra/extracellular and whole-body communication channels. Moreover, their inherent magnetism facilitates easy detection by external devices, positioning SPIONs as a potential bridge for interfacing information between nanonetworks and these external systems.




    \section{Calcium ions (\text{Ca}${2+}$)}

    Calcium signaling is a pivotal cellular mechanism within the body, playing a vital role in a variety of biological processes. Specifically, it is essential for muscle contraction in smooth muscle cells, modulates neuronal activities in astrocytes, and underpins pancreatic functioning in exocrine cells\cite{berridge2000versatility,petersen2008polarized}. Because of its profound significance, it is extensively investigated by molecular biologists and biophysicists, leading to the establishment of numerous detailed models\cite{berridge2003calcium, rudiger2014stochastic, graef1999type}.

    At the cellular level, the propagation of Ca$^{2+}$ ions is more than just a series of chemical events—it is a sophisticated communication process, which has been studied under the paradigm of molecular communication\cite{nakano2005molecular}. This angle not only provides a novel perspective on this vital biological process but also highlights its potential adaptability for communication between nanomachines. In recent research, communication channel models centered on calcium signaling have been formulated\cite{bicen2016linear, barros2015comparative}, and the feasibility of deploying Ca$^{2+}$-based communication has been conceptualized \cite{kuran2012calcium}. By combining with network theory, communication theory metrics tailored for calcium-based communication networks provide invaluable insights, especially when examining the repercussions of conditions such as Alzheimer's Disease\cite{barros2018multi}. Here,  we explore the intricate dynamics of calcium signaling, its implications for molecular communication, and how these insights could reshape our understanding of both biological processes and emerging technologies.

    \subsection{Physical Characteristics}
    Ca$^{2+}$ ions are present both in the extracellular fluid and within the cytosol of cells. The concentration of Ca$^{2+}$ is instructive for various biological processes. Several cellular organelles and membrane transport systems closely regulate these concentrations. Intracellular concentrations of Ca$^{2+}$ are typically much lower than extracellular concentrations. This concentration gradient is actively maintained by membrane transport systems such as the plasma membrane Ca$^{2+}$-ATPase (PMCA) and the sodium-calcium exchanger (NCX), both of which function to pump Ca$^{2+}$ out of the cell.
    
    Upon specific cellular stimuli, the endoplasmic reticulum (ER), a primary intracellular storage site for Ca$^{2+}$, releases these ions into the cytosol, leading to an increase in intracellular calcium concentration. This release acts as a signaling event, instigating various biological responses\cite{simons1988calcium}. Importantly, Ca$^{2+}$ ions do not directly facilitate long-distance information transfer between cells. Instead, a localized increase in Ca$^{2+}$ concentration in one cell can trigger the release of Ca$^{2+}$ in neighboring cells. This relayed mode of stimulation ensures the propagation of the signaling event across multiple cells, allowing for coordinated cellular responses\cite{bicen2016linear}.

    \begin{enumerate}
        \item 
        \textit{Types ---}
        The Ca$^{2+}$ ion, by definition, has only one possible ionic structure, which implies that type-based information encoding isn't feasible. It is noteworthy that Ca$^{2+}$ exists naturally in the human body, underscoring its biocompatibility. Given its small size, mass, and general biocompatibility, Ca$^{2+}$ emerges as a promising candidate for in-body cell-level nanodevice communication.
        
        \item
        \textit{Mass ---}
        The mass of Ca$^{2+}$ is approximately 40.078 atomic mass units (u) which can be calculated accurately from the atomic components of the ion. This value can be precisely determined from the atomic components of the ion.

        \item 
        \textit{Charge and Magnetic Susceptibility ---}
        The calcium ion lost two electrons, making them to be positively charged with +2e. this positive charge makes them subject to attraction from negatively charged entities, which may affect the molecular communication channel. For instance, anions in solution, negatively charged regions on molecules (like the carboxyl groups on certain amino acid residues), and cellular structures like the inner face of the cell membrane which can be negatively charged due to the phospholipid composition\cite{ma2017introducing} which will affect the propagation of the Ca$^{2+}$.

        \item 
        \textit{Dimensions and Diffusivity ---}
        Calcium ions (Ca$^{2+}$) are often depicted as spherical entities with a radius of approximately 0.1 nanometers\footnote{It is important to note that the atomic radius is not a fixed value; the boundary of an atom is not sharply defined due to the diffuse nature of the electron cloud.}. These nanoscale dimensions enable Ca$^{2+}$ ions to reside within cells and propagate through ion channels, facilitating the relay system's functionality.
        The diffusivity of Ca$^{2+}$ in the cytoplasm is measured to be 5.3$\times$10$^{-6}$ cm$^2$s$^{-1}$ as reported in \cite{donahue1987free}. Because of the small size and therefore high diffusivity, Ca$^{2+}$-based molecular communication channel can have a relatively high data rate in free diffusion environments.

        \item 
        \textit{Biocompatibility ---}
        As previously noted,  Ca$^{2+}$ions are prevalent in the human body due to their pivotal roles in various biological functions. Consequently, they are inherently biocompatible, making them suitable for in-body application scenarios.

    \end{enumerate}

    \subsection{Communication Channel and Techniques}
    
    As introduced earlier in this section, calcium signaling is intricate and multifaceted. Rather than facilitating a straightforward transport of Ca$^{2+}$ over distances to convey information, the process leverages various cellular organelles and operates in a relay fashion. A study in \cite{berridge2003calcium} offers a thorough review of the biophysical complexes of this mechanism. However, given the diversity and specificity of mechanisms across different biological systems, the details can become too complicated, especially for an initial channel modeling. To address this, a more streamlined version of the model from \cite{bicen2016linear} is presented here, which captures the main structure of the calcium signaling systems. It is anticipated that this model will serve as a foundational platform, with further refined channel models being developed by incorporating specific characteristics pertinent to each system under consideration.

    \begin{figure}[ht]
     \centering
     \begin{subfigure}[b]{.38\textwidth}
         \centering
         \includegraphics[width=\textwidth]{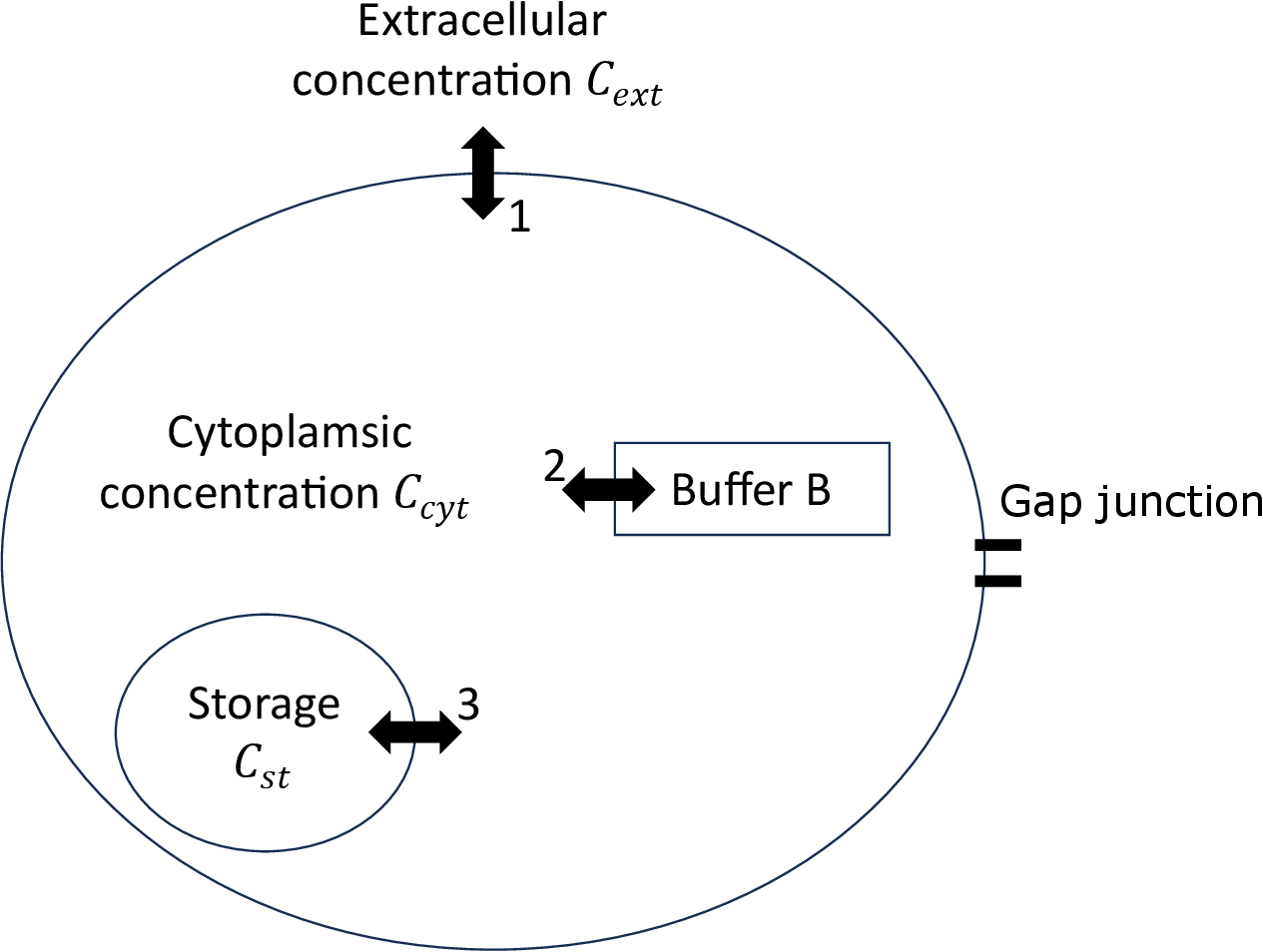}
         \caption{}
         \label{fig: Ca generator}
     \end{subfigure}
     \hfill
     \begin{subfigure}[b]{.5\textwidth}
         \centering
         \includegraphics[width=\textwidth]{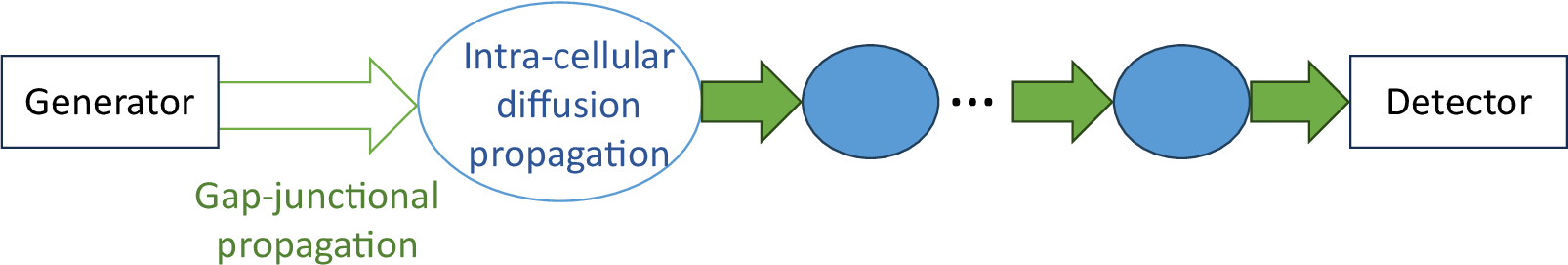}
         \caption{}
         \label{fig: relay}
     \end{subfigure}
    \caption{(a) The diagram depicts the dynamics of calcium signaling within this cell, which are governed by the interplay of extracellular ($C_{ext}$), cytoplasmic ($C_{cyt}$), and storage ($C_{st}$) calcium concentrations, modulated by buffer B. (b) Illustration of the relay propagation channel for the Ca$^{2+}$ signal wave, highlighting two primary mechanisms: gap-junctional propagation for intercellular communication, and intracellular diffusion propagation for signal spread within a cell.
}
    \label{fig: Ca channel}
    \end{figure}

    As shown in Figure \ref{fig: Ca channel} (a), the Ca$^{2+}$ signal wave is generated by a generator cell, which is a dynamical system that contains three variables extracellular concentration $C_{ext}$, cytoplasmic concentration $C_{cyt}$ and storage concentration $C_{st}$. The buffer organelles are made up of molecules that can bind with Ca$^{2+}$ and thereby dampen the Ca$^{2+}$ change, such as calbindin\cite{blatow2003ca2+} and calretinin\cite{schwaller2014calretinin}. The store organelles are ER and Mitochondria\cite{kuran2012calcium}. Other than the parameter of buffer B, there are a number of other parameters that govern the dynamics of the generator cell. For a full demonstration of the generator model, refer to\cite{bicen2016linear, berridge2003calcium}. To produce a Ca$^{2+}$ signal, $C_{ext}$ is increased, which will lead to a change in $C_{cyt}$ according to the differential equations governing the generator cell. Then, the change of $C_{cyt}$ will propagate through intracellular cytosol and ion gates to reach the next cell that would repeat the mechanism of generating the Ca$^{2+}$ and pass on the signal. Effectively, these will form a relay of calcium signals over a channel made up of a block of cells. One linear channel in the block of cells is shown in Figure \ref{fig: Ca channel} (b).

    This biophysical model of the calcium propagation channel has made a number of simplifications. Firstly, between the generator cell and the detector cell, there is usually a block of cells, which makes the transmission channel more than a linear channel.
    Secondly, there will be regenerations of Ca$^{2+}$ in the mediating cells due to the propagation of IP$_3$ together with the Ca$^{2+}$ and the PLC$\delta$ activity, which will induce Ca$^{2+}$ release from local storage in the cells\cite{kuran2012calcium}. This is called the Calcium-induced calcium release (CICR). This effect is neglected in this model. Thirdly, the calcium wave may also travel extracellularly to transfer information. Finally, the specific propagation mechanism will be different due to differences in biological elements in different cells. For example, the different mechanisms in smooth muscle cells, epithelial cells and astrocytes are studied separately in \cite{barros2015comparative}
     Still, this linear channel model captures the framework of calcium signaling, which could be used for ICT metric analysis in the next subsection. Moreover, the model could be further extended if it is studied for a specific cellular system.

    \begin{enumerate}
        \item
        \textit{Transmitter ---}
        For the provisioning of Ca$^{2+}$ in cellular processes, organisms rely on natural reserves. An average human requires a daily intake of 1,000 to 1,500 milligrams of calcium to uphold typical calcium blood levels without compromising bone integrity\cite{us2019calcium}. This calcium is mainly stored in the bones and certain intracellular compartments.
    
        In the calcium ion propagation model in Figure \ref{fig: Ca channel}, the initiation of calcium signaling is dependent on an increase in the extracellular concentration of Ca$^{2+}$. In nature, this signaling can be triggered through ligand binding with proteins, or via electrical stimulation\cite{tsien1983calcium}. For experimental or artificial inductions, an ion pump can be utilized to surge the Ca$^{2+}$ concentration, effectively acting as a transmission mechanism\cite{isaksson2007electronic}. Additionally, the creation of synthetically engineered Ca$^{2+}$ wave generator cells presents another potential method\cite{pham2011synthetic}. However, the miniaturization of these tools to be compatible with nanomachines remains a pivotal challenge, a necessary step to translate nanonetwork concepts into practical applications.

        \item
        \textit{Propagation Channel ---}
         As shown in Figure \ref{fig: Ca channel} (b), the propagation channel of the Calcium signal has a relay structure, where the cells between the generator and the detector act as the relay media. The intercellular propagation is accomplished by gap junctions (structures that exist in the plasma membranes of adjacent cells that allow the exchange of various ions, second messengers, and small metabolites). The propagation in a gap junction is a diffusion process subjected to the permeability of the gap junction and the number of open gap junctions. On the other hand, intracellular propagation is modeled as a free diffusion process.
        
         Unlike free diffusion, the specialized structure of the propagation channel (blocks of cells) requires extra effort to establish. It is proposed in \cite{kuran2012calcium} that other than a preconfigured channel, the transmitter and receiver could be designed to form intermediate cells by themselves to construct the propagation channel. However, this approach relies on advanced bioengineering techniques, and there does not appear to be dedicated research addressing this problem.
         
        \item
        \textit{Receiver ---}
        The receiver is considered a point-wise detector that could measure the concentration of the Ca$^{2+}$ in the current model. A precise measurement of the concentration of Ca$^{2+}$ is a challenge. In nature, the influx or change in Ca$^{2+}$ concentrations can trigger conformational changes in proteins, subsequently modulating their functional activities\cite{bagur2017intracellular}. For example, the fluorescence signal of Ca$^{2+}$ sensitive dyes\cite{zanin2019methods} or microelectrodes that penetrate the cell membrane\cite{hove2010making} can be used for measuring Ca$^{2+}$ concentration. However, Ca$^{2+}$ sensitive dyes may be subject to interference from other elements in the channel such as Mg$^{2+}$ and the detection of fluorescence requires a specialized cumbersome setup, while microelectrodes also require a specialized setup and can only be applied to larger cells. The measurement of Ca$^{2+}$ at the nanoscale is still a great challenge and requires further research.


        \item
        \textit{Modulation ---}
        In nature, biological activities are governed by the presence or absence of Ca$^{2+}$ ions. Effectively, it can be stated that information is encoded in Ca$^{2+}$ waves using On-Off Keying (OOK) modulation. Since there is only one type of Ca$^{2+}$ ion, type-based modulation is not applicable. Additionally, the propagation of Ca$^{2+}$ ions depends on local release, meaning that the spatial distribution is not preserved. Consequently, spatial-based modulation is also not applicable. However, by accurately parameterizing the communication channel and techniques, time-based modulation could be adapted to enhance the data rate of the communication system \cite{barros2014transmission}.

    \end{enumerate}




        
        

    \subsection{Communication Performance}
    \begin{enumerate}
        
        \item 
        \textit{Data rate ---}
        Due to the requirement of a cell channel, calcium signaling is used for microscale communication. In astrocytes, the maximum propagation range of Ca$^{2+}$ wave is 200-350$\mu$m in radius and has a velocity of 15-27$\mu$m/s\cite{kang2009spatiotemporal}. In \cite{barros2015comparative}, the channel models of calcium signaling for astrocytes, epithelium cells, and smooth muscle cells are compared, where epithelium reaches a capacity of 0.01 bit/s with an interference probability of about 0.1\% for a distance of 1 cell.

        \item 
        \textit{Noise sources ---}
        A primary source of error in Ca$^{2+}$-based molecular communication is ISI. ISI arises from residual Ca$^{2+}$ ions left behind from previous diffusion events, which can overlap with subsequent signals and distort the intended message. Additionally, recurrent noise, resulting from Ca$^{2+}$ ion waves reflecting off cellular boundaries, poses challenges for signal clarity\cite{barros2014transmission}.

        Another complicating factor is extracellular interactions. The presence of other divalent cations, such as Mg$^{2+}$, can compete with Ca$^{2+}$, potentially interfering with its detection or signaling functionality\cite{jing2018many}. Moreover, the communication environment is inherently complex. Rather than a simplistic linear channel, it more accurately resembles a network or block of cells. In such a scenario, Ca$^{2+}$ signals originating from different sources might interfere with one another.
    
        Furthermore, several biological processes unrelated to the desired communication could inadvertently release Ca$^{2+}$ ions, adding an element of unpredictability to the system. It is also essential to consider the inherent noise associated with the generation and detection of calcium waves.


    \subsection{Communication Applications}
    Calcium signaling plays a pivotal role in cellular processes, and the calcium-ion-based communication channel modeling has been aimed at exploring this mechanism from an ICT perspective and for using them as the information molecule for general nanonetwork communication at the cellular level since the calcium ions mainly work on the microscale in intracellular and extracellular space. Considering this natural process within the molecular communication paradigm offers a fresh perspective for its analysis. This approach has the potential to inspire novel diagnostic and therapeutic applications for diseases related to calcium dysregulation\cite{barros2018multi}. With the advent of artificial cells, coupled with advancements in miniaturized and precise calcium wave generators and detectors, calcium-based communication offers promising avenues for facilitating interactions between nanomachines and cells, especially in systems that already exist naturally within the human, such as the muscular and nervous tissues\cite{barros2017ca2+}.

    \begin{figure}
        \centering
        \includegraphics[width = 2.7in]{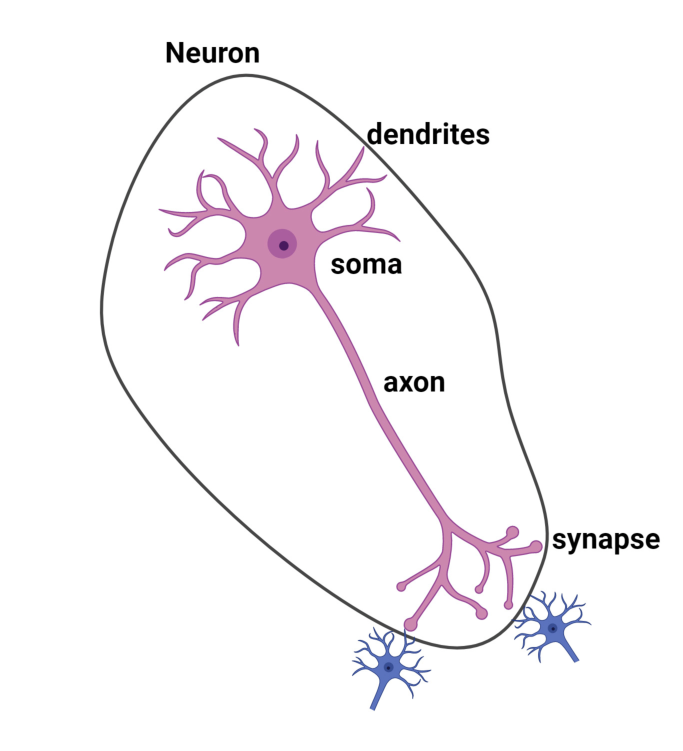}
        \caption{The schematic illustrates a neuron's key components: dendrites, soma (cell body), axons, and synapses. Signals from the previous neuron are transferred from the dendrites to the soma, then travel down the axon, cross the synapses, and finally reach the dendrites of the next neurons. Synaptic communication occurs between the synapse and the dendrites.}
        \label{fig: Neuron}
    \end{figure}

    \end{enumerate}

    \section{Neurotransmitters}
    The neurotransmitter serves as a key molecule for transmitting information between neurons, which are the fundamental units of the brain and nervous system. These neurons play a pivotal role in receiving, transmitting, and processing information from sensory and motor organs to the brain. Structurally, a neuron comprises three components: the soma, axon, and dendrites, as illustrated in Figure \ref{fig: Neuron}. Information transmission between neurons occurs through two modes: electrical and chemical. Electrical transmission is facilitated via axonal signal transmission, whereby an action potential, or a voltage change, diffuses along the axons, sequentially triggering the opening of ion gates \cite{betts2020anatomy}. Conversely, chemical transmission occurs at the junctions between the axon termini and the dendrites of adjacent neurons, where synaptic communication takes place. During this process, the presynaptic terminal releases neurotransmitters that traverse the synaptic cleft (the gap between the presynaptic and postsynaptic terminals) to reach the receptors on the postsynaptic terminals. Upon binding to these receptors, neurotransmitters could induce an action potential that further propagates along subsequent axons, thus continuing the information relay. This process of synaptic transmission aligns with the paradigm of molecular communication, wherein the presynaptic terminal, synaptic cleft, and postsynaptic terminal function as the transmitter, propagation channel, and receptor, respectively. Accordingly, several models have been explored\cite{ramezani2017communication}, some accounting for various channel properties, such as plasticity \cite{khan2019impact, malak2013synaptic}, astrocyte effects and extracellular matrix interactions \cite{lotter2020synaptic}, as well as spillover and reuptake in the channel \cite{veletic2019synaptic}. Through such investigations, researchers aim to obtain a deeper understanding of critical parameters within the synaptic transmission system, such as the receptor binding rate and the depolarization threshold. This knowledge could prove instrumental in comprehending neurological pathologies and enhancing the diagnosis and treatment thereof. Additionally, developing a comprehensive model of the synaptic transmission system is fundamental for designing devices, like brain-machine interfaces with synaptic stimulator electrodes, which require precise control over individual neurons \cite{veletic2019synaptic}.


  \begin{figure}[htp]
        \centering
        \includegraphics[width=3.5in]{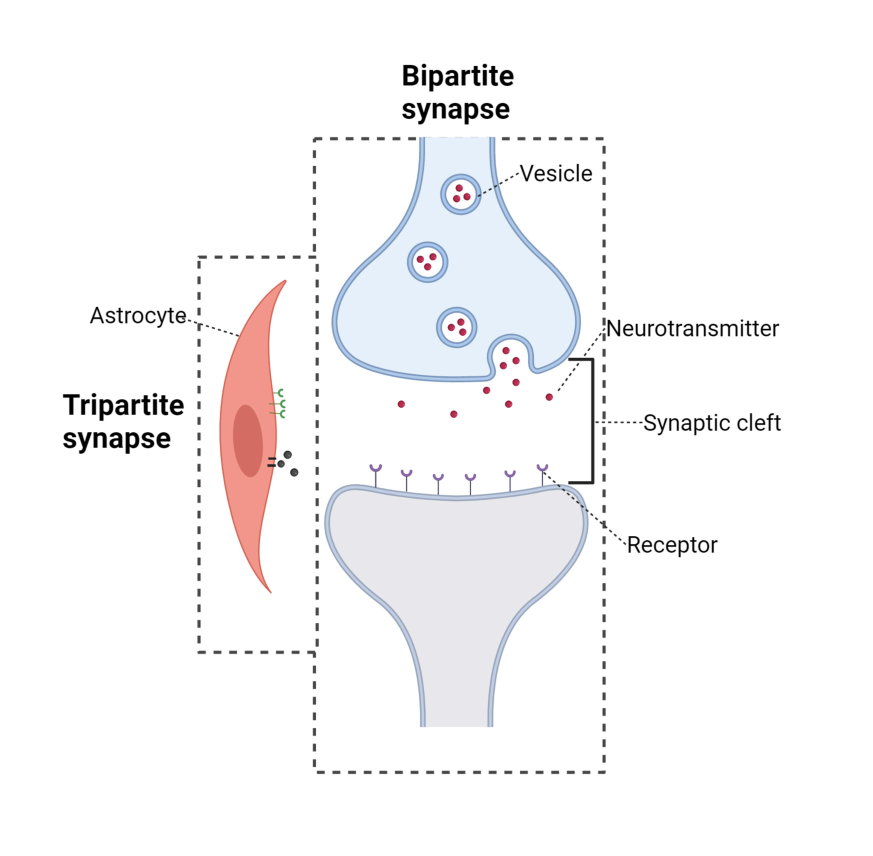}
        \caption{This schematic diagram illustrates two configurations of synaptic communication systems. The bipartite synapse, enclosed within the dotted box on the right, entails the vesicles transmission from the presynaptic terminal to the postsynaptic terminal by diffusing in synaptic cleft. The tripartite synapse model, represented in the full image, incorporates astrocytes adjacent to the synaptic components.}
        \label{fig: synapse}
        \end{figure}

        \subsection{Physical Characteristics}
        \begin{enumerate}

            \item
            \textit{Type ---}
            Many neurotransmitters exist in nature, and historically, defining them has been challenging\cite{hyman2005neurotransmitters} because of the biological complexity of synaptic communication. Typically, the neurotransmitters can be classified into three categories: small-molecule neurotransmitters, peptide neurotransmitters, and other neurotransmitters based on their chemical composition. There are many types of neurotransmitters of different functionality and physical properties contained in each category. For instance, \textit{substance P} is a peptide neurotransmitter in the spinal cord that could help suppress pain, while Nitric oxide (NO), being gaseous, is considered a type of other neurotransmitter that could mediate the synaptic plasticity\cite{Libretexts_2022}. Due to these diverse properties, the synaptic communication channel for each neurotransmitter must be characterized respectively.
            Currently, the neurotransmitter most extensively researched under the paradigm of molecular communication is \textit{glutamate}\cite{lotter2020synaptic,khan2017diffusion, veletic2015communication, veletic2016peer}, a type of small-molecule neurotransmitter. As the primary excitatory neurotransmitter in the brain\cite{meldrum2000glutamate}, its biophysical mechanism is relatively well-understood, with studies dating back to the 1950s\cite{hayashi1952physiological}. Here, ``excitatory'' indicates that when glutamate binds to its receptor, the firing of an action potential on the postsynaptic terminal becomes more probable. Conversely, gamma-aminobutyric acid (GABA) serves as the primary inhibitory neurotransmitter. Its inhibitory effect, when combined with glutamate's excitatory effect, affects the probability of action potential firing. \cite{chapman2022yin}. The binding of different neurotransmitters has cumulative effects on neuronal firing. Therefore, introducing a new neurotransmitter into the system will not merely add an extra encoding dimension, as seen in type-based modulation. In natural systems, such introductions are often associated with more intricate biological mechanisms like synaptic plasticity\cite{chapman2022yin} and the overall balance of neuronal activity in the brain.
            
            \item
            \textit{Mass ---}
            As discussed earlier, different neurotransmitters possess distinct masses, determined by their chemical composition. For instance, glutamate, with the chemical formula $\text{C}_5\text{H}_9\text{NO}_4$ has a molar mass of 147.15u (or g/mol). This translates to a molecular mass of approximately  2.44$\times$10$^{-25}$ kg for each glutamate molecule. Similarly, a GABA molecule weighs around 1.71$\times$10$^{-25}$ kg. These minuscule masses enable neurotransmitters to rapidly diffuse across the synaptic cleft, facilitating swift neuronal communication and responses.

            \item
            \textit{Charge and Magnetic Susceptibility ---}
            Neurotransmitters can exhibit varying charges based on their chemical composition; for instance, at physiological pH, glutamate carries a negative charge\cite{barnes2003bioinformatics}, GABA acts as a zwitterion (a molecule exhibiting both positive and negative charges yet remains electrically neutral)\cite{Roberts:2007}, and dopamine is positively charged\cite{liu2021biosensors}. It is noted that the local electric field due to ion current from ion gates could reach $10^4$ V/m, therefore affecting the mobility of the charged neurotransmitters\cite{rusakov2011shaping, sylantyev2008electric}. Methods like Transcranial Direct Current Stimulation and Deep Brain Stimulation utilize non-invasive electrodes to direct current to the brain, treating conditions like depression\cite{nitsche2009treatment} and enhancing motor and cognitive performance\cite{thair2017transcranial}. However, these methods primarily influence the membrane potential rather than the propagation of neurotransmitters. Moreover, the direct impact of magnetic susceptibility on neurotransmitter propagation is not well-documented in scientific literature. While the influence of magnetic fields on ion channels and membrane potential has been explored\cite{bertagna2021effects}, their direct contribution to neurotransmitter dynamics pales in comparison to the significance of the neurotransmitter type and concentration, which will be further elaborated upon in the following discussion. 

            \begin{figure}
                \centering
                \includegraphics[width=2.6in]{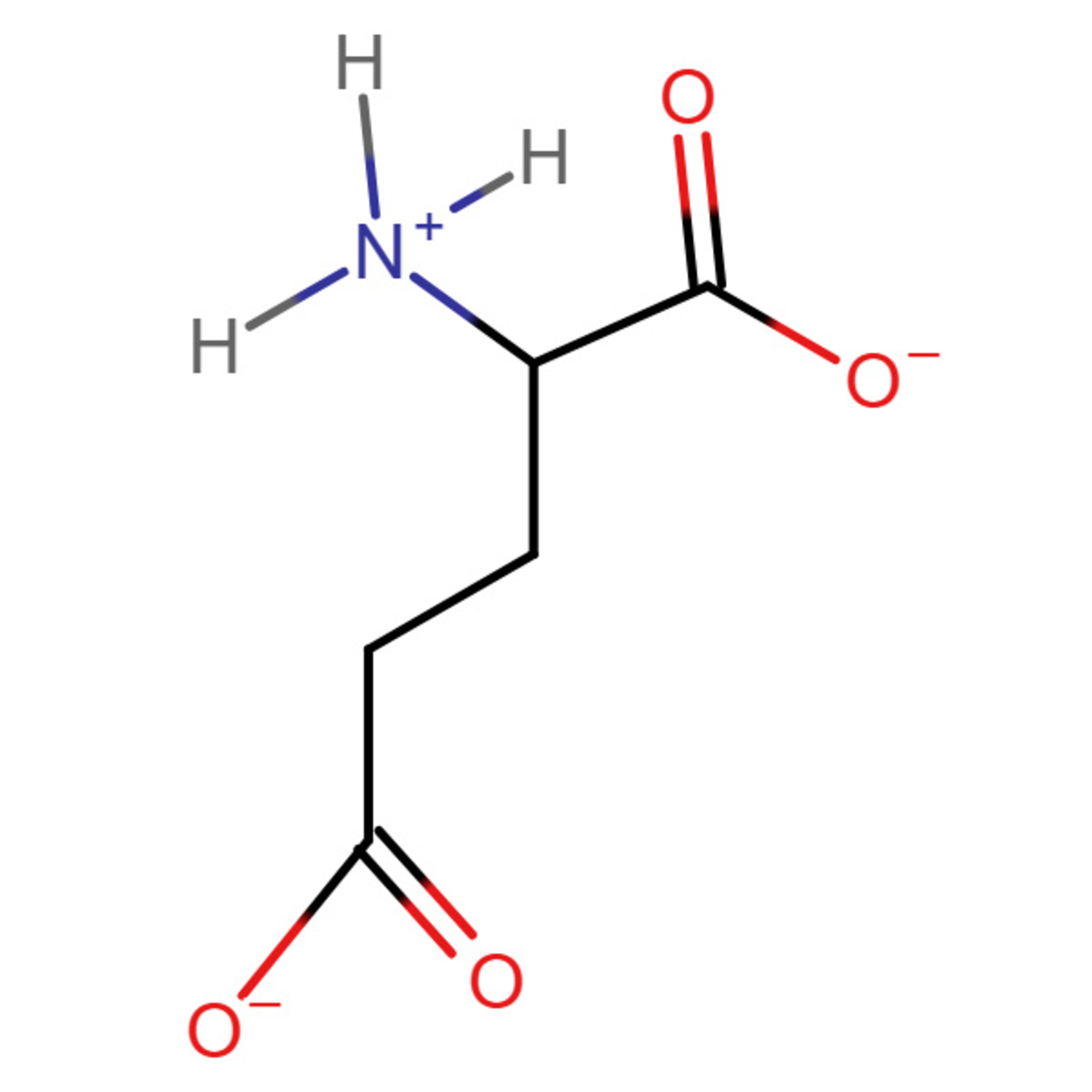}
                \caption{The chemical structural formula of a glutamate\cite{pubchem_2019b}.}
                \label{fig: glutamate}
            \end{figure}

            \item 
            \textit{Dimensions and Diffusivity ---}
            As previously mentioned, neurotransmitters can be categorized broadly into two types based on size: \textit{small molecule transmitters} and \textit{neuropeptides}. The former category includes transmitters like GABA and glutamate, which are individual amino acids. They are typically stored in vesicles that are between 40 to 60 nm in diameter. On the other hand, neuropeptides, comprised of sequences ranging from 3 to 36 amino acids, usually are stored in vesicles of 90 to 250 nm\cite{neuroscience2001}. The sizes of the small molecule transmitters can be approximately deduced from their chemical structure as shown in Figure \ref{fig: size comparison}. For instance, as shown in Figure \ref{fig: glutamate}, the backbone of glutamate is made up of four carbon-carbon bonds, connected with two carboxy. Therefore, to give a rough estimation of the lengths of the bonds. The length of the C-C bonds is 0.154 nm\cite{harris2015bond}, and the length of the carbon-oxygen bonds is also around 0.1nm. Therefore, taking the 3D shape of the glutamate into account, a very rough estimation of the size of the glutamate can be obtained to be $\sim$ 0.6 nm.

            This distinction in size has implications: it can influence both the diffusion coefficient and the rate at which these neurotransmitters bind to receptors. As such, size is a crucial parameter in communication models that account for these properties. Further exploration into the variations among neurotransmitters will be discussed in the subsequent subsection.
            
            To illustrate the point about diffusivity, a study cited in \cite{nielsen2004modulation} reported that the diffusivity of glutamate in the synaptic cleft fluctuates between 0.25 to 0.42 $\times$ 10$^{-8}$ cm$^2$/s, averaging out at 0.33 $\mu$m$^2$/ms. It is vital to note, however, that there is a wide variety of neurotransmitters, and the specific properties of the synaptic cleft can also influence diffusivity. For instance, the same study found that the presence of Dextran, a macromolecule, could decrease the diffusivity to as low as 0.17 m$^2$/ms\cite{nielsen2004modulation}. In comparison, somatostatin-14 (SST), a type of neuropeptide, is measured to have $~$0.09 $\mu$m$^2$/ms in tissue slices\cite{xiong2021probing}. This highlights the necessity for realistic channel models of synaptic communication to individually assess both the neurotransmitters involved and the unique properties of their propagation channels.

        \end{enumerate}

\begin{table*}
    \centering
    \caption{Summary of Biological Mechanisms in the Synaptic Communication System}
    \label{tab: synaptic channel}
    \begin{tabular}{>{\centering\arraybackslash}m{3cm}|>{\centering\arraybackslash}m{8cm}|>{\centering\arraybackslash}m{3cm}}
    \hline
    \textbf{Communication} \newline \textbf{Component}  & 
    \textbf{Biological Mechanisms} & 
    \textbf{References}  \\ \hline
    \noalign{\vskip 2mm} 
    \hline
    \textit{Transmitter}\newline(Presynaptic Terminal) &
    \begin{itemize}[leftmargin=*]
        \item Linear-Nonlinear-Poisson (LNP) Model
            \begin{itemize}
                \item Axonal transmission
                \item Saturation of presynaptic neurons
                \item Stochasticity of presynaptic spike train
            \end{itemize}
        \item Pool-based vesicle model
        \item Stochastic vesicle release
        \item Astrocyte feedback
    \end{itemize}
     &
    \cite{malak2013communication}, \cite{simoncelli2004characterization}, \cite{rizzoli2005synaptic}, \cite{zhang2015improved}, \cite{veletic2015communication} \\ 
    \hline
    \textit{Propagation Channel} (Synaptic Cleft) &
    \begin{itemize}[leftmargin=*]
        \item Three Dimensional rectangular cuboid boundary
        \item Brownian motion (Diffusion)
        \item Spillover to ECM
        \item Glial uptake
        \item Presynaptic reuptake
        \item Enzymatic degradation
        \item Noise from other neurons
    \end{itemize}
     &
    \cite{veletic2019synaptic}, \cite{lotter2020synaptic}, \cite{jamali2019channel}, \cite{neuroscience2001}, \cite{anderson2000astrocyte} \\ 
    \hline
    \textit{Receiver}\newline(Postsynaptic Terminal) &
    \begin{itemize}[leftmargin=*]
        \item Ligand-Receptor Binding
        \item Different types of receptors and alpha functions
        \item EPSP generation and AP generation
        \item Reversible/irreversible adsorption
        \item Saturation of receptors
        \item Lateral diffusion of receptors (affect distribution of receptors)
    \end{itemize}
     &
    \cite{veletic2019synaptic}, \cite{rao2007nmda}, \cite{bigharaz2016realistic}, \cite{lotter2020synaptic}, \cite{rusakov2011shaping}, \cite{miyazaki2021excitatory} \\
    \hline
    \end{tabular}
\end{table*}

        \begin{figure}[htp]
        \centering
        \includegraphics[width=3in]{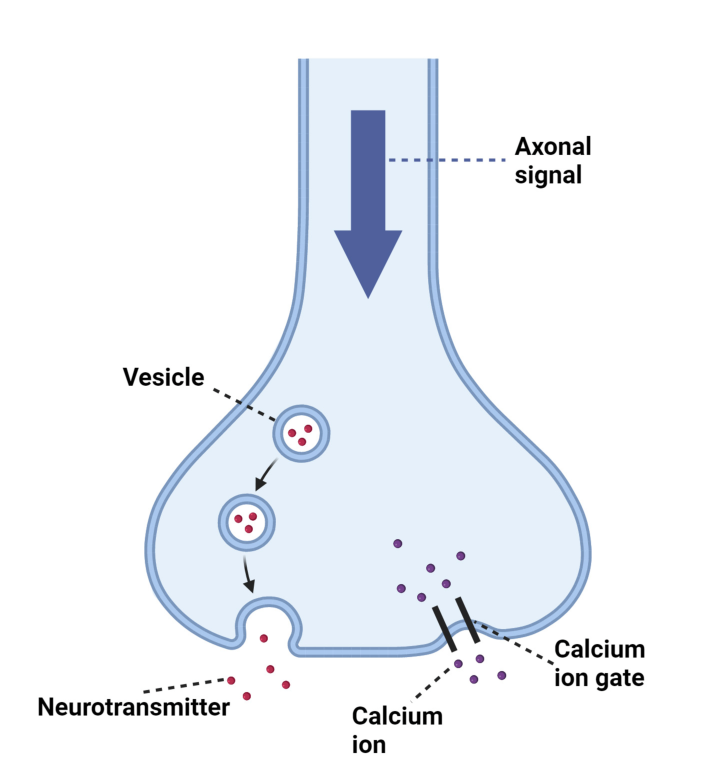}
        \caption{This schematic diagram depicts the presynaptic terminal of a neuron during synaptic transmission. An axonal signal triggers the opening of calcium ion gates. The influx of calcium ions causes neurotransmitter-filled vesicles to fuse with the membrane and release their contents into the synaptic cleft. This sequence of events, critical for the propagation of neural signals, is further explained in the main text.}
        \label{fig: presynaptic terminal}
        \end{figure}

        \subsection{Communication Channel and Techniques}
        Synaptic communication is a complicated process, with multiple biological mechanisms occurring within each component of the system. These processes together facilitate the sophisticated functions of neural networks. In this communication framework, the presynaptic terminal serves as the transmitter, the synaptic cleft as the propagation channel, and the postsynaptic terminal as the receiver. Within the synaptic cleft, neurotransmitters primarily propagate through diffusion. However, their movement and interactions are also influenced by various other mechanisms, which are summarized in Table \ref{tab: synaptic channel}. We will delve into in this section with reference to the established models\cite{lotter2020synaptic, khan2017diffusion, veletic2016peer,balevi2013physical, ramezani2018sum, malak2014communication,malak2013communication}.

        \begin{enumerate}
            \item
            \textit{Transmitter ---} 
            The presynaptic terminal acts as the transmitter in the synaptic communication system. As illustrated in Figure \ref{fig: presynaptic terminal}, the procedure of neurotransmitter release unfolds in the following manner: Firstly, the action potential (AP) — an electrical signal generated by the soma to transmit information along the axons — reaches the presynaptic terminal. This causes ion gates to open for Ca$^{2+}$ ions. Secondly, the influx of Ca$^{2+}$ prompts vesicles, which are membrane-bound sacs containing neurotransmitters, to fuse with the presynaptic terminal's membrane\cite{sudhof2012calcium}. Finally, as these vesicles fuse, they release neurotransmitters into the synaptic cleft. The inherent biological mechanisms and their stochasticity have led to the proposal of multiple mathematical models. As described in \cite{malak2013communication}, the axonal transmission is represented by a low-pass filter. Along with the saturation of the neurons (which occurs when readily releasable vesicles are depleted), and the stochasticity of the presynaptic spike train (represented by both a point nonlinearity model and a Poisson model), a Linear-Nonlinear Model is constructed to depict the signal reaching the presynaptic terminal. Furthermore, a pool-based, stochastic model is employed to describe the fusion of the transmitter. In this model, vesicles are categorized into three pools\cite{malak2013communication}: the Readily Releasable Pool (RRP)\cite{ramezani2017information}, the Recycling Pool, and the Reserved Pool, as illustrated in Figure \ref{fig: pool-based model}. The RRP primarily releases neurotransmitters. In the Recycling Pool, vesicles undergo endocytosis and are refilled with neurotransmitters. The Reserved Pool serves as a larger backup reservoir. Key parameters, such as the pool sizes and refilling rates, significantly influence the transmitter's overall performance. Moreover, in a neural network, multiple neurons may be connected to the same neuron, making it necessary to have a multiple-access channel model to make the model more realistic\cite{malak2013communication}. Additionally, feedback from astrocytes may further affect the presynaptic terminal, a mechanism explored in the subsequent section on the propagation channel\cite{veletic2015communication}.

            \begin{figure}[htp]
            \centering
            \includegraphics[width=2.2in]{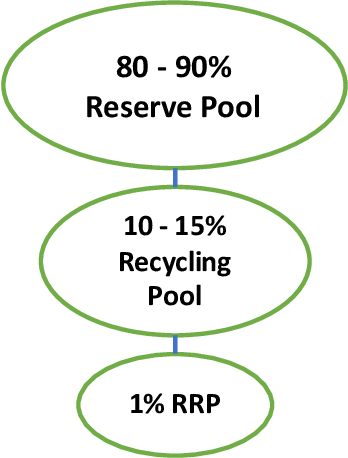}
            \caption{This diagram illustrates the proportional distribution of vesicle pools in the presynaptic terminal. The Readily Releasable Pool (RRP), which contains vesicles that are immediately available for neurotransmitter release, only takes up 1\% of the pool.}
            \label{fig: pool-based model}
            \end{figure}

            \item 
            \textit{Propagation Channel ---}
            In the synaptic communication system, the synaptic cleft serves as the propagation channel. In \cite{lotter2020synaptic}, the channel is modeled as a 3-dimensional rectangular cuboid, where different boundaries represent various biological processes occurring within. The movement of neurotransmitters within the synaptic cleft is characterized as a diffusion-reaction system. While this movement is primarily governed by the diffusion equation, attributing to the Brownian motion of the neurotransmitters, several other reactions also play crucial roles. Firstly, a phenomenon known as \textit{spillover} occurs, where neurotransmitters overflow into the extracellular matrix and may influence neighboring neurons \cite{veletic2019synaptic}. Secondly, astrocytes can actively \textit{uptake} neurotransmitters to remove excess from the synaptic cleft, reducing the ISI and facilitating high-frequency neuronal communication \cite{anderson2000astrocyte}. The consideration of the astrocytes elevates the bipartite system (pre/postsynpatic terminal) to a tripartite system as shown in Figure \ref{fig: synapse}. Additionally, the presynaptic terminal may reuptake neurotransmitters present in the synaptic cleft. This reuptake process is represented by a boundary condition of the third kind in \cite{lotter2020synaptic}. Intriguingly, rather than merely uptaking the neurotransmitter, astrocytes, once triggered by the uptaken glutamate, might release glutamate that reciprocally affects both the presynaptic and postsynaptic terminals \cite{veletic2015communication}. Lastly, there is \textit{enzymatic degradation}. In this process, specific enzymes break down the neurotransmitters present in the cleft\cite{neuroscience2001}. However, the impact of enzymatic degradation on the communication channel is considered minor compared to glial uptake and is thus overlooked in \cite{lotter2020synaptic}.

            \item 
            \textit{Receiver ---}
            In the communication channel, the postsynaptic terminal serves as the receiver. More precisely, the receptors on the cell membrane of the postsynaptic terminal act as the actual receivers. For the information to be relayed to the next neuron, the receiving neuron must fire an action potential based on the neurotransmitters captured by these receptors, which is detailed below.

            Neurotransmitter receptors fall into two primary categories:\textit{ ionotropic} and \textit{metabotropic}. Ionotropic receptors, upon binding with neurotransmitters, undergo an immediate conformational alteration. This change either opens or closes an associated ion channel, permitting ion flow. In contrast, metabotropic receptors do not directly control ion channels. Instead, they initiate intracellular signaling cascades by leveraging G-proteins, indirectly influencing the channel's activity\cite{betts2020anatomy}. Consequently, their response is slower. Of the glutamate ionotropic receptors, AMPA and NMDA are predominant. AMPA receptors have a direct response to glutamate by opening ion channels, facilitating rapid reactions, and are commonly employed in swift synaptic transmissions. NMDA receptors, conversely, need postsynaptic polarization (potentially from AMPA receptor activation) to become active. They are more resistant to activation, but once opened, their ion channels remain active longer than AMPA channels, making them crucial for synaptic plasticity\cite{rao2007nmda}. When neurotransmitters are received, these receptors open ion channels, inducing an Excitatory PostSynaptic Potential (EPSP). The receptors' response to glutamate can be represented using alpha functions\cite{bigharaz2016realistic}. If the cumulative EPSP surpasses a set threshold, the postsynaptic neuron fires an action potential. This impulse then travels down the axon of the succeeding neuron, perpetuating information transmission.
            
            Several factors influence this reception process. Firstly, the number of receptors on the postsynaptic terminals is finite, and they can transition between unbound and bound states as they interact with neurotransmitters. Should neurotransmitter concentration become exceedingly high, these receptors might saturate, causing the EPSP to peak. Secondly, for instances of irreversible binding (where neurotransmitters cannot detach back into the synaptic cleft), the ligand-receptor binding model is apt\cite{kuscu2019channel}. However, for reversible binding, this model falls short\cite{lotter2020synaptic}. Thirdly, while receptors are typically viewed as being uniformly scattered across the postsynaptic terminal\cite{lotter2020synaptic}, mechanisms such as lateral diffusion can alter receptor distribution, directing neurotransmitters to otherwise unreachable areas\cite{rusakov2011shaping}. Lastly, while prevailing models prioritize excitatory receptors, a comprehensive representation should also incorporate inhibitory receptors, like the GABA receptors, to achieve enhanced realism\cite{miyazaki2021excitatory}.

            \item \textit{Modulation ---}
            The modulation inherent in synaptic communication differs from that in previous molecular communication systems, primarily because it integrates with axonal transmission driven by the propagation of Action Potentials. In essence, both the concentration and types of neurotransmitters (whether excitatory or inhibitory) are utilized for OOK modulation where the firing and non-firing of the action potential correspond to 1 and 0.
            
        \end{enumerate}    
        \subsection{Communication Performance}
        \begin{enumerate}
        
        \item 
        \textit{Noise Source ---}
        Several factors introduce noise in the synaptic communication channel:

        \begin{itemize}
        \item       
        \textit{Thermal noise ---}
        Just as electronic devices experience thermal noise due to the random motion of electrons, biological systems can experience noise due to the random motion of ions and molecules, especially at the synaptic cleft where neurotransmitters diffuse across the small space\cite{stevens1972inferences}.    
        \item 
        \textit{Probabilistic Vesicle Release ---}
        The release of neurotransmitters from vesicles is inherently probabilistic. Even under consistent conditions, not every vesicle releases its content upon receiving an action potential\cite{branco2008local}.
        \item 
        \textit{Stochastic Ligand-Receptor Binding ---}
        The binding process between neurotransmitters and their receptors is also probabilistic, introducing another source of variability.
        \item 
        \textit{Uptake Variability ---}
        The uptake of neurotransmitters by glial cells, as well as presynaptic reuptake, adds further randomness to the process.
        \item 
        \textit{Background Neuronal Activity ---}
        Neurons typically receive inputs from numerous other neurons, and this concurrent background activity introduces additional noise.
        \end{itemize} 
        
        Despite these sources of variability, the synaptic Communication Channel and Techniques can handle high-frequency neural signal transmissions due to mechanisms like astrocyte uptake, presynaptic reuptake, and enzymatic degradation, which collectively minimize the ISI.
        
        While individual synaptic channels might be noisy, this noise can paradoxically be advantageous when viewed in the context of larger collective neural networks. Such noise can enhance the propagation of information\cite{destexhe2022noise} and bolster the network's information processing capabilities\cite{guo2018functional}. Mechanisms like stochastic resonance can transform this noise into a functional tool, amplifying weak signals and making them more detectable\cite{mcnamara1989theory, longtin1993stochastic}. Moreover, the inherent variability from synaptic noise can prevent the system from becoming overly deterministic, allowing for more adaptive and flexible responses, and possibly facilitating phenomena like exploration-exploitation trade-offs in neural computations

        \item 
        \textit{Data rate ---} Experimentally, in \cite{abbasi2018controlled} an in-vivo nervous communication channel is built using earthworms, achieving a data rate of 66.6 bit/s with a BER of 6.8 $\times$ 10$^{-3}$. Furthermore, in \cite{veletic2016upper}, the estimated upper bound for the data rate of a bipartite system, using parameters derived from realistic measurements, stands at 1.6 bit/s. In contrast, \cite{de1996rate} reports a significantly higher data rate of approximately 50 bit/s. This data pertains to information transmission through chemical synapses to large monopolar cells (LMCs) in blowflies. One plausible explanation for this substantial difference in data rates might lie in the source of the parameters used in the studies. The parameters employed in \cite{veletic2016upper} is based on measurements from the rat hippocampus, as detailed in \cite{kang1998astrocyte}. Naturally, these would differ from those of blowfly neurons. Nonetheless, to bridge such discrepancies and gain a deeper understanding, there is a pressing need to develop more realistic models. These models should encompass a wider range of biological mechanisms. Moreover, directly comparing these models with experimental data will be crucial to validate their accuracy and relevance.

        \end{enumerate}
        
        \subsection{Communication Applications}
        Therefore, synaptic communication operates on the nanoscale, with synaptic clefts typically spanning a length of 20-50nm\cite{gabbiani2017mathematics, cox2010synaptic}. Owing to its nanoscale and dependency on the biological infrastructure (pre/postsynaptic terminals), the primary applications of synaptic communication are for understanding and improving synaptic communication, i.e., nervous signaling channels. Gaining a comprehensive understanding of the communication theory behind synaptic systems offers profound insights. For instance, it elucidates the malfunctioning synaptic communication observed in diseases like Alzheimer's, Parkinson's\cite{lepeta2016synaptopathies,taoufik2018synaptic}, and brain cancer\cite{monje2020synaptic}. Furthermore, this knowledge holds promise in developing diagnostic methods centered around molecular communication model parameters. Beyond disease understanding and diagnostics, synaptic communication plays a key role in fundamental brain functions like learning and information processing, intricately tied to synaptic plasticity. Lastly, with advancements in artificial synapses\cite{yu2021evolution}, there is a vision of crafting neural prostheses to either replace or supplement defective synapses. Similarly, the potential for synaptic-modulating brain-machine interfaces emphasizes the importance of a robust comprehension of molecular communication using neurotransmitters\cite{veletic2019synaptic}.

    \section{Odor molecules}
    The odor molecules include a broad range of molecules that could propagate in air and liquid and be detected by natural or artificial sensors. For example, alcohol, acetone (the smell of nail polish), and green leaf volatiles (the smell of grass) are all odor molecules. These odor molecules are often considered principal information carriers for macro-scale molecular communication. Another important example is \textit{pheromones}, typically secreted by animals and plants, which can propagate through the air to trigger specific behaviors. For instance, ants use them for food foraging guidance, and plants use them to signal distress during an attack. This natural communication system aligns perfectly with the molecular communication paradigm. Investigating the information-theoretical aspects of these molecules can provide deeper insight into this crucial biological mechanism in animals and plants \cite{unluturk2016end}. Moreover, this communication method has wide potential applications, such as in swarm robotics for urban waste management \cite{alfeo2019urban}, due to its low energy consumption, durability, and effectiveness in challenging environments with obstacles \cite{purnamadjaja2010bi, kube2000cooperative}. Other envisioned applications may include agricultural monitoring, underwater and in-mine communications, and diagnosis of diseases related to the olfactory bulb \cite{Dilara2023odor}.

        \subsection{Physical Characteristics}
       Despite the various types of odor molecules, they share many similar physical characteristics while still being different in many ways. Moreover, the odor molecules, in the context of molecular communication, are considered for different scenarios. Artificial odorants such as alcohol are often used for building proof-of-concept experimental testbeds, while pheromones are often considered in the context of natural communication between animals and plants.
        \begin{enumerate}
            
            \item \textit{Charge and Magnetic Susceptibility ---}
            Odor molecules are predominantly neutral because their lack of electric charge makes them less susceptible to electrostatic interactions, thereby increasing their volatility and enabling them to propagate in the air. Examples include methanol and ethanal (types of alcohol), jasmonic acid (a plant pheromone), and bombykol (a moth pheromone), all of which are electrically neutral. However, the chemical nature of these molecules can change depending on the environment, including factors such as the solvent, pH value, and temperature. For instance, acetic acid, a weak acid, can release a proton in solution to form acetate ions. In aquatic environments, specific sulfur compounds, such as hydrogen sulfide (H2S), can undergo oxidation to form sulfate ions ($SO_4^{2-}$) under certain conditions, particularly in the presence of oxidizing agents\cite{millero1987oxidation}. Therefore, environmental factors impact on the electrical charge of molecules must be carefully considered when investigating their channel models. 
            Regarding their magnetic susceptibility, odor molecules are primarily organic and exhibit diamagnetism due to their electronic configurations\cite{broersma1949magnetic}. This characteristic means they are slightly repelled by a magnetic field. However, the magnetic field's influence on an odor-molecule-based communication system is minimal in practical scenarios(with molar magnetic susceptibility on the order of $\approx$10$^{-6}$). In contrast, magnetized ferrofluids have been proposed as an alternative to pheromones for swarm robot communication, as outlined in \cite{brenes2022magnetic}. These ferrofluids share essential properties with pheromones, such as locality, diffusion, and evaporation. Additionally, their magnetic nature facilitates the detection of their presence and concentration, providing an advantage over the less sensitive artificial detectors designed for other odor molecules. Nonetheless, while these ferrofluids can effectively leave a trail for robot navigation, they are incapable of propagating through the air, preventing them from establishing an air-based molecular communication channel.

            \item \textit{Types ---} There are reportedly over 400,000 types of odor molecules\cite{kuroda2023human, bonner1950odeurs}. In nature, individual scents are often composed of combinations of these odor molecules, theoretically creating an almost infinite number of complex odors. Humans can distinguish more than a trillion different olfactory stimuli, utilizing about 500-750 different kinds of odorant receptors \cite{neuroscience2001}. This remarkable perceptual capacity makes type-based modulation especially suitable for communication channels based on odor molecules. Given the vast array of odor molecules, nature has developed specialized detection methods. These natural strategies can be adapted to create modulation methods for artificial odor-based communication channels\cite{jamali2023olfaction}, details of which will be elaborated in the following sections.

            \item
            \textit{Mass ---}
            To ensure high volatility, odor molecules typically possess minimal mass. According to \cite{turin2003structure, ohloff1994scent}, the heaviest known odorant is a type of labdane, with a molecular weight of approximately 296 Daltons, approximately 4.25 $\times$ 10$^{-25}$ kg. The high volatility allows the odor molecules to propagate in the air, making them more suitable information molecules for macroscale communication over the molecules in the previous sections.

            \item \textit{Diffusivity ---} 
            First, the sizes of the odor molecules are very small, typically smaller than a few nanometers because it is hard for molecules bigger than 20 carbons (about 3 nm) to diffuse effectively\cite{purnamadjaja2010bi,bradbury1998principles}. For example, the moth pheromones are hydrocarbon chains, of 10-18 carbons in length\cite{resh2009encyclopedia}, ethanols (alcohol) are 0.4 to 0.5 nm and aldehy around 0.1 nm (the length of its CO bond)\cite{turin2003structure}.
            This small size is crucial for the molecules' ability to propagate efficiently in the air, enabling long-range molecular communication. For instance, the diffusion coefficients of common alcohols in the air at room temperature are approximately 0.1 - 0.2 cm$^2$/s \cite{lapuerta2014equation}, facilitating their rapid dispersal and enhancing their effectiveness as signaling molecules.
            
            \item \textit{Biocompatibility ---} Odor molecules are primarily considered for macroscale communication in the air or liquid, rather than for in-body communication. For pheromones and odorants, they disperse in the open air and are detected by an organism's odor receptors, triggering a subsequent reaction. This means they are not introduced directly into the body, such as into vessels or cells. As for alcohols, they are mainly utilized in proof-of-concept experiments, so biocompatibility is not a primary concern. However, certain alcohols can be non-toxic and may be used in blood vessels for therapeutic purposes, provided their concentration remains below safety thresholds\cite{orlando2014ethanol}.
        \end{enumerate}

        \begin{figure}
            \centering
            \includegraphics[width = 3.3in]{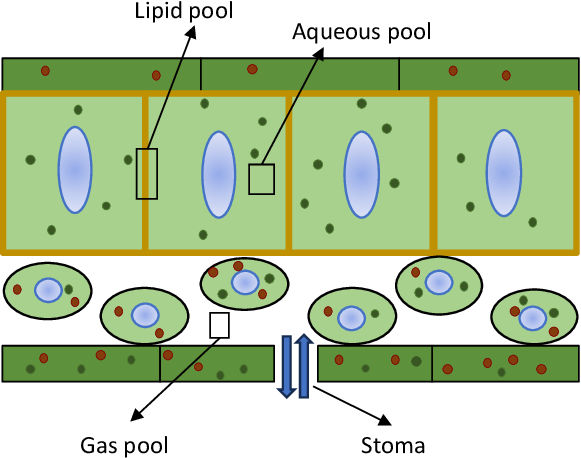}
            \caption{This is a simplified structural diagram of leaves. There are mainly three pools of pheromones: lipid, aqueous, and gas pool. The pheromone and other particles, such as oxygen and carbon dioxide, are exchanged with the surroundings through the stoma. The diagram is adapted from \cite{unluturk2016end}, and a more detailed 3D illustration of the leaf anatomy can be found at \cite{Mukherjee_2022}.}
            \label{fig: plant transmitter}
        \end{figure}

        \begin{figure}
            \centering
            \includegraphics[width=3.3in]{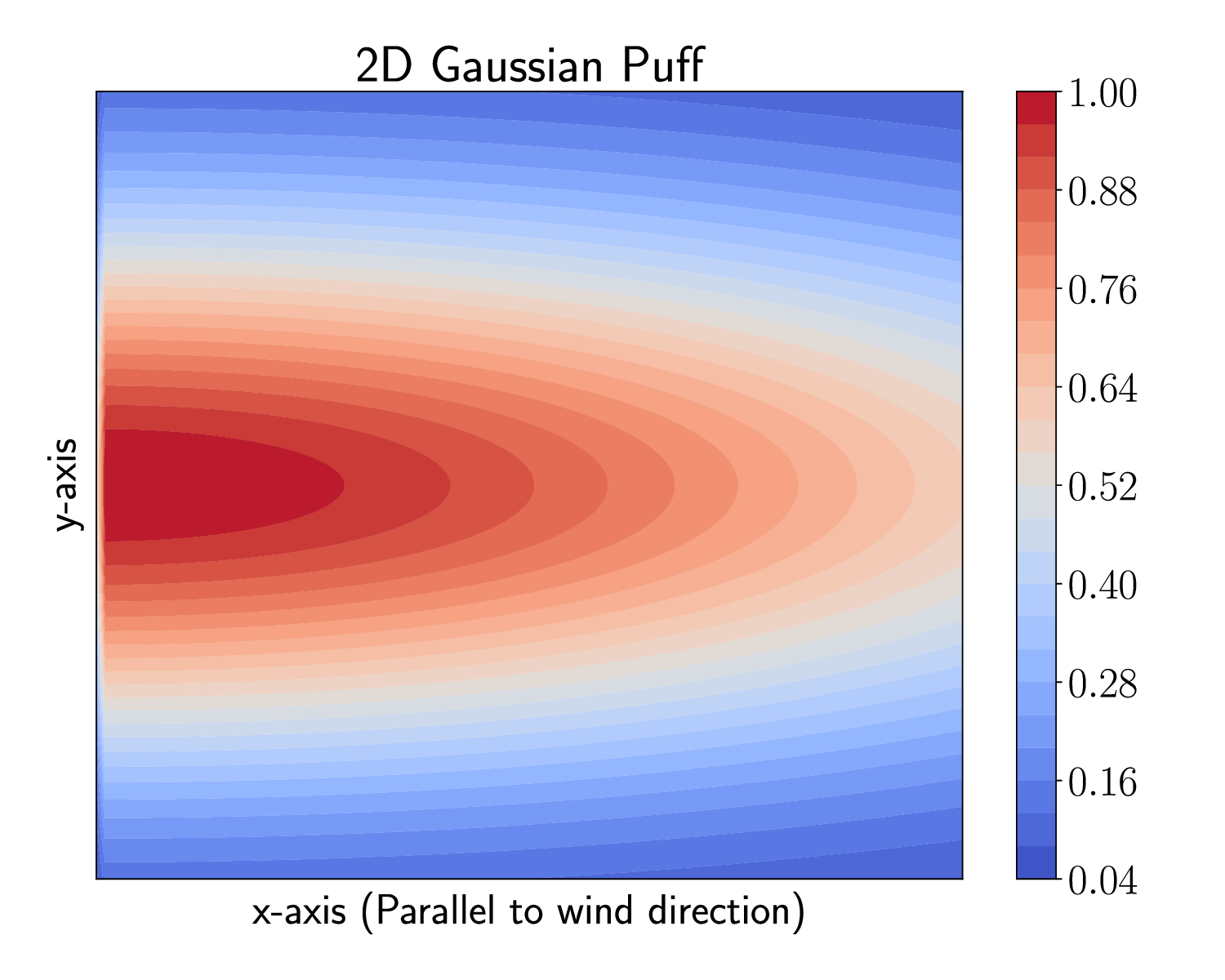}
            \caption{This figure demonstrates the shape of a Gaussian puff, representing the diffusion of odor molecules in the air influenced by the wind. }
            \label{fig: gaussian puff}
        \end{figure}
        \subsection{Communication Channel and Techniques}

        \begin{enumerate}
            \item \textit{Transmitter---} In \cite{unluturk2016end}, the natural transmitter structure of plants is explored, detailing how pheromones are stored in different phases within various parts of the leaves, as depicted in Figure \ref{fig: plant transmitter}. The biophysical model of the dynamics between the pools and exchange with the surroundings through the stoma is explored in \cite{niinemets2002model} which is effectively governed by a set of differential equations. Then, the model is further integrated with the propagation and receiver model for an end-to-end channel model of plant communication in \cite{unluturk2016end}.
            
            In contrast, the emission process of odor molecules in artificial communication systems can be considerably more straightforward. Typically, it is envisaged that these odor molecules are synthesized externally and housed within a storage unit at the transmitter. Electric pumps can then facilitate the emission process \cite{purnamadjaja2010bi}. For simplification, the transmitter may be modeled as a point source, as discussed in previous sections. Nonetheless, a realistic transmitter emits odor molecules with a distinct spatial and temporal distribution, necessitating a comprehensive model for accurate representation. An instance of this complexity is found in \cite{purnamadjaja2010bi}, where the angular distribution of pheromones used in robot communication is measured and analyzed, underscoring the need for a detailed approach to transmitter modeling.

            \item \textit{Propagation Channel ---} The propagation of odor molecules, both in air and liquid, can be modeled by considering diffusion, advection (the flow of air or liquid), and turbulence\cite{unluturk2016end, jamali2019channel}. Typically, in the direction of the flow, the advection will be dominant, whereas, in the transverse direction, free diffusion will take effect, leading to what is often characterized as a Gaussian puff\cite{zannetti1990gaussian}, as illustrated in Figure \ref{fig: gaussian puff}. Turbulence is factored into the model by adjusting the diffusion coefficient to account for eddy diffusivity\cite{unluturk2016end}. Regarding the boundary conditions, the macro-scale range of the envisioned communication channel often classifies it as an unbounded channel. However, specific scenarios, such as communication within pipes, necessitate a bounded model. Moreover, the communication channel will not resemble a simplistic three-dimensional rectangular cuboid, as seen in synaptic communication. Instead, more complex channel geometries, potentially including numerous obstacles, must be considered in modeling the channel, especially in environments such as urban areas or mines.

            \item \textit{Receiver ---} In nature, the reception of odor molecules is a complex process that varies among different species and types of odor molecules \cite{leal2013odorant}. The biophysical models of these receptions are not fully established. In this discussion, we qualitatively introduce the odor molecule reception processes in human noses \cite{sharma2019sense} and in general, plant leaves \cite{unluturk2016end}. Subsequently, we explore some possibilities for the detection devices of artificial odor molecules.

            For humans, the detection process begins when odor molecules inhaled into the nasal cavity travel through the mucus layer to reach the Odorant-Binding Proteins (OBPs) in the Olfactory Epithelium. The OBPs transport them to the odorant receptors (ORs). These receptors identify molecules matching their specificities and relay signals to the olfactory bulb, pre-processing them before forwarding them to the brain. It is important to note that this is a simplified overview; a more comprehensive description is available in \cite{sharma2019sense}. For species like moths, additional components, such as antennae, play a crucial role in detection \cite{leal2005pheromone}.
            
            In plants, the reception of pheromones can be somewhat simplified as a diffusion process through a ``doorway,'' considering that pheromones inside the leaves might reside there, not contributing to the concentration gradient necessary for diffusion \cite{trapp1995generic, unluturk2016end}.
            
            In the field of artificial odor molecule detection, several relevant technologies exist. Initially, metal oxide semiconductors were widely utilized as detectors due to their sensitivity to gases through a tin dioxide (SnO$_2$) film \cite{farsad2013tabletop, lu2016vertical}. The presence of these gases triggers a reaction with the tin dioxide, changing the resistance of the material  which can be captured by the circuit. E-noses employ arrays of metal oxide sensors, reacting to various odor molecules in a mixture and creating a unique fingerprint based on the array's collective reaction pattern \cite{li2014recent}. When combined with machine learning techniques, e-noses exhibit considerable proficiency in identifying different compounds within a mixture \cite{ye2021recent}. For more precise analysis, Gas Chromatography-Mass Spectrometry (GC-MS) systems are effective. These systems separate compounds in a mixture based on the principle that various compounds interact differently with materials and travel at distinct speeds. The separated compounds are then analyzed by mass spectrometry to discern their characteristics by their mass-charge ratio. However, the GC-MS system, with its large size, non-portability, high cost, and lengthy processing time, requires further refinement for broader practical application scenarios. Alternatively, conducting polymer sensors and quartz crystal microbalance sensors present viable options \cite{purnamadjaja2010bi, li2014recent}. While these detectors are more affordable and convenient, they lack the sensitivity offered by GC-MS systems. In macroscale molecular communication, without the requirement of miniaturization, the discussed detectors can in effect be implemented directly as receivers in odor-molecule-based molecular communication systems.
            
            \item \textit{Modulation ---} As previously mentioned, concentration-based and type-based modulations are particularly suitable for odor-molecule-based molecular communication. Moreover, in nature, odor receptors can respond to different molecules with varying intensities. By interpreting the combined reactions of different receptors, organisms can distinguish more odors than the number of receptors they possess. Mimicking this mechanism, a method known as \textit{Molecule Mixture Shift Keying} has been proposed for molecular communication, with the potential to significantly enhance communication performance with capable hardware \cite{jamali2023olfaction}.

        \end{enumerate}

        
        \subsection{Communication Performance}
        \begin{enumerate}
            \item \textit{Noise Sources ---} In molecular communication, noise sources on the transmitter side primarily stem from limitations in the design and manufacturing of the transmitter itself. For instance, the released molecules do not emanate from an ideal isotropic point source. In reality, their distribution fluctuates both angularly and temporally, increasing the unpredictability of the channel. Within the propagation channel, various factors contribute to noise, including background interference from other odor molecules in the open air, intersymbol interference from previous signal transmissions, uncontrollable turbulent flow in air or liquid, and molecular degradation. Some of these challenges can be mitigated by establishing the communication channel in a more controlled environment, and utilizing clean, regulated air or liquid flows to clear the channel before transmission, thereby reducing intersymbol interference (ISI). However, in broad application scenarios, such as urban swarm robotics or underwater communication, these complications persist. On the receiver side, noise may arise from the stochastic firing of neurons, the thermal noise inherent in electronic detectors, and the difficulty of accurately distinguishing different types of odor molecules. To diminish noise and enhance channel performance, the implementation of more sophisticated detectors and error-correcting coding methods is needed.

            \item \textit{Data Rate ---} Numerous testbeds utilizing alcohol and acid/base have been established due to their ease of production and detection. For instance, \cite{koo2016molecular} demonstrated a MIMO archetype of alcohol-based communication, achieving a bit rate of 0.34 bit/s and a bit error rate of 9.75$\times$10$^{-2}$. The experiment in\cite{lu2017vertical} explored a vertical alcohol communication channel, leveraging gravity to attain a rate of 2 bit/s over a distance of 0.05 - 0.1m. Meanwhile, \cite{farsad2017novel} introduced a system where acid and base molecules were transmitted, using pH value detection for information decoding. This approach realized a data rate of 0.3 bit/s in an open-air environment. It is pertinent to note, however, that these are preliminary proof-of-concept experimental testbeds employing rudimentary experimental devices. It is anticipated that future iterations will witness the construction of higher-performance channels, facilitated by improved modulation, enhanced detectors, and a broader spectrum of odor molecules.
            
        \end{enumerate}

        \subsection{Communication Applications}
        As a macro-scale communication method, odor-molecule-based communication systems stand poised to complement traditional electromagnetic (EM) communication methods, especially in challenging environments such as underground \cite{sun2009underground} and underwater \cite{mcguiness2019experimental, lanbo2008prospects}. On the other hand, by deciphering the intricacies of odor communication systems within the body, we can significantly advance the diagnosis and treatment of diseases related to the olfactory system \cite{Dilara2023odor}. Moreover, pheromones are at the center of plant and animal communication\cite{shorey2013animal} such as fungi communication\cite{cottier2012communication} and moths\cite{allison2016pheromone}. Therefore, inspecting pheromone communication from the perspective of molecular communication can lead to application in the agriculture and food industry such as optimizing plant growth environment and monitoring pest invasions. Currently, pheromones serve merely as signaling molecules for navigation in swarm robotics. However, with continued development, odor-molecule-based communication systems hold the promise of achieving higher information capacity. This advancement is crucial as it underscores their potential role in long-duration information retention, akin to an information mailbox for swarm robots, enhancing their collaborative functionalities. A comprehensive discussion of the applications of odor-molecule-based communication can be found at \cite{Dilara2023odor}.

    \section{Other molecules}
    This section briefly introduces other various information molecules that are explored with the molecular communication paradigm. Generally, these molecules share common attributes with the previously mentioned entities, such as biocompatibility, nanoscale dimensions, and stability. However, each possesses distinct characteristics, making them suitable for diverse scenarios. The array of molecules discussed encompasses quantum dots, organic dyes, aerosols, RNAs, other ions, vesicles, proteins, Phosphopeptides, sugars, polystyrene microbeads, cAMPs, AHLs, and IPTGs. This section is intended to encompass all the molecules investigated in this field to date, serving as a navigational aid for those seeking an in-depth understanding of this research area.
    \subsection{Quantum Dots}
\begin{figure}
        \centering
        \includegraphics[width = 3.5in]{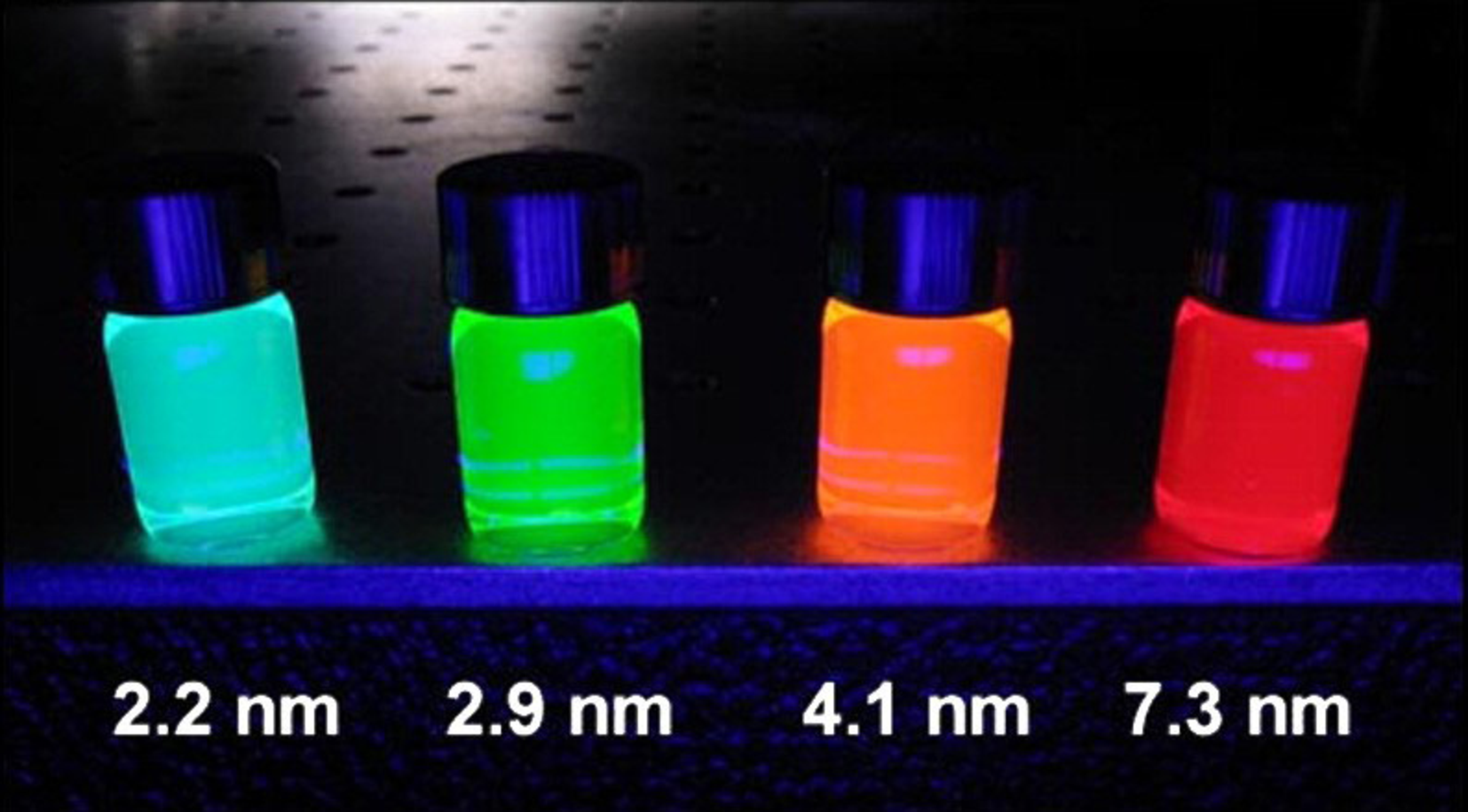}
        \caption{A photo of quantum dots of different sizes. The colors of the quantum dots are determined by their sizes. The figure is cited from \cite{smith2008bioconjugated}}
        \label{fig: quantum dots}
    \end{figure}

    Quantum dots, semiconductor nanoparticles, exhibit unique electronic and optical properties that depend on their size and composition. Specifically, adjusting their size and structure allows them to produce fluorescence at specific wavelengths when excited by a UV light source, as demonstrated in figure \ref{fig: quantum dots}. Moreover, quantum dots are relatively straightforward to synthesize, offering precise control over their properties. These characteristics have led to significant applications in fields such as medical imaging\cite{smith2006multicolor} and the display industry\cite{liu2020micro}, culminating in the awarding of the Nobel Prize in Chemistry in 2023 for their discovery and synthesis.

    Among the various chemical compositions of quantum dots, carbon quantum dots are considered information molecules due to their biocompatibility\cite{lim2015carbon}, effective dispersion and diffusion in fluids owing to their nanosize\cite{cali2022effect}, ease of preparation, and relatively low cost\cite{lim2015carbon}. Additionally, their presence can be optically detected\cite{cali2022effect}. These artificial particles are envisaged for use in communication between nanomachines, predominantly employing diffusion-based propagation channels in liquid mediums (potentially with advection or turbulence). Current modulation methods focus on the concentration of quantum dots (i.e., the intensity of the fluorescence) at the terminal\cite{cali2022fluorescent, cali2022effect}, but more sophisticated detection systems might enable type-based modulation, where detectors identify a mixture of quantum dots based on the emitted fluorescence wavelengths.
    
    In terms of applications, quantum dots could operate in scenarios similar to those of magnetic nanoparticles, spanning both meso and microscales. At the mesoscale, experimental testbeds visible to the human eye could validate quantum-dot-based molecular communication and implantable medical devices are anticipated to function within this range\cite{cali2022fluorescent}. At the microscale, researchers have established experimental testbeds using quantum dots to study the effect of Taylor diffusion (a phenomenon where solute molecules experience enhanced diffusivity in a flow system) on molecular communication channel performance\cite{cali2022effect}.
    
\begin{figure}
        \centering
        \includegraphics[width = 3.5in]{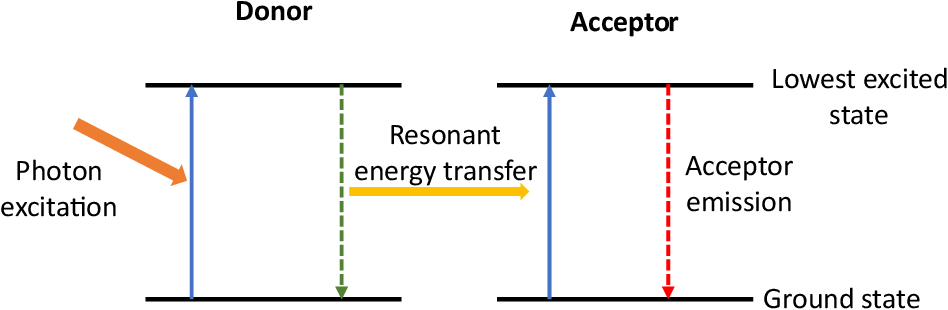}
        \caption{The figure shows the process of FRET. The donor electron is excited by a photon in this example. Its energy is then transferred to the acceptor by the non-radiative resonance process and excites the electron in the acceptor. The acceptor electron then deexcites by emission of a photon which can be detected by a photon sensor.}
        \label{fig: FRET}
    \end{figure}

    Beyond free diffusion, Forster Resonance Energy Transfer (FRET) represents another potential propagation channel for information in quantum dots. FRET, occurring among fluorophores like quantum dots\cite{dos2020quantum} and fluorescent proteins\cite{bajar2016guide}, involves the spontaneous transfer of energy from donor fluorophores to nearby acceptor fluorophores with similar electronic properties as shown in Figure \ref{fig: FRET}. This energy transfer can serve as a means of communication, offering advantages over diffusion channels\cite{kuscu2011physical} such as a faster data transfer rate and lower environmental dependency, thus providing higher controllability. Although a single information transfer via FRET is limited to a very short range (approximately 10nm), an information relay system comprising a network of nanomachines capable of transferring information through FRET can overcome this limitation\cite{kuscu2014communication, kuscu2015internet}.

    \subsection{Organic Dyes}
    Organic fluorescent dyes, similar in use to quantum dots, are instrumental in building experimental testbeds for molecular communication, as demonstrated in \cite{abbaszadeh2019mutual}. An example is provided in \cite{damrath2021investigation}, which describes the development of an air-based molecular communication channel using Uranine and Rhodamine 6G. This innovative approach incorporates a camera-based detector and achieves a data transmission rate of 40 bit/s across a span of 2 meters, utilizing modulation methods dependent on both concentration and type. However, it is pertinent to note that many organic dyes, including Rhodamine 6G, carry significant toxicity \cite{alford2009toxicity}, precluding their use in applications within the human body. Furthermore, the FRET-based communication of fluorescent dyes is investigated in \cite{solarczyk2016nanocommunication}, with multiple donors and acceptors.
    In parallel, color pigments have been chosen to create proof-of-concept platforms for molecular communication due to their affordability and the ease with which they can be detected \cite{pan2022molecular}. 
    
    \subsection{Aerosols (with Pathogens)}
    Aerosols, suspensions of fine solid particles or liquid droplets in gas, are pivotal in the airborne transmission of diseases. These pathogens, expelled through coughing, sneezing, laughing, or exhaling, hitch a ride on aerosols, enabling their spread through the air. This sequence—coughing (transmission), airborne aerosol movement (propagation), and reaching another individual (detection)—alongside the transmission of pathogens in the human respiratory system, falls under the molecular communication paradigm \cite{koca2021molecular}.
        The COVID-19 pandemic has cast a spotlight on aerosol-based transmission. These particles, typically less than 5 micrometers in diameter \cite{koca2021molecular}, are larger than the molecules previously discussed but smaller than respiratory droplets, allowing them to remain suspended in the air for prolonged periods. Consequently, their propagation channel is chiefly diffusion with potential advection or turbulence in open air \cite{khalid2020modeling}. The modulation method resembles OOK, where infection presence constitutes a 1 and its absence a 0. Given its macro-scale nature, successful modeling of aerosol-based communication \cite{gulec2021molecular, thakker2022modelling, hoeher2021mutual} can yield insights into disease transmission dynamics, exemplified by studies on coronavirus transmission through air and within the human respiratory system \cite{schurwanz2021duality,koca2021molecular}
.
    \subsection{RNAs}
    Similar to DNA, RNA is composed of a series of nucleotide bases. However, it differs from DNA in functionality and is considered a key molecule in cell-to-cell communication\cite{chen2022plant}. RNA is made up of a different set of nucleotides: A, G, C, U, and is typically single-stranded with a complex 3D geometry. Unlike DNA, which serves as a stable store of genetic information, RNA plays active roles, such as transcribing DNA, allowing it to interact directly with other biological components like ribosomes for protein synthesis. There are various types of RNA, each serving distinct functions. For example, messenger RNA (mRNA) carries the genetic code from DNA to the ribosome, serving as a template for protein synthesis, while ribosomal RNA (rRNA), in conjunction with proteins, forms the structural components of ribosomes, the cellular machines that synthesize proteins. Therefore, the molecular communication channel models of RNA require specific consideration for different RNA types in various scenarios, an area currently underexplored.

    \subsection{Other Ions}
    Beyond the calcium ions discussed in the previous section, researchers have explored various other ions as information molecules in molecular communication. These include hydrogen ions (H$^+$) and hydroxide ions (OH$^-$), which influence acidity\cite{farsad2017novel, grebenstein2019molecular, walter2023real}, and sodium and chloride ions, components of table salt \cite{wang2020understanding, angerbauer2023salinity}. These ions are selected primarily because both they and their respective detectors (for salinity and pH) are readily accessible. Upon dissolution in water, their propagation is governed by diffusion, potentially influenced by advection or turbulence. Experimental testbeds employing these ions serve as proof-of-concept prototypes for molecular communication systems and offer opportunities to explore strategies for enhancing channel performance. For instance, \cite{farsad2017novel} applies a machine learning algorithm to establish detection thresholds, thereby improving channel performance. Similarly, \cite{grebenstein2019molecular} describes an experimental testbed reliant on engineered bacteria designed to emit protons, with environmental pH levels serving as the information medium.
    On the other hand, a number of different other ions are used as information carriers in various communication systems from different contexts. In \cite{lin2020adaptive}, the lithium ions are effectively working in a communication channel between the electrodes of the resistive random access memory (RRAM) device, which mimics the pre-and post-synaptic terminals of a neuron and could be used for neuromorphic computing. Other ions, including sodium and potassium, are fundamental to various biological processes, such as nerve impulses, muscle contractions, and heart rhythms\cite{soderlund2010sodium,mackinnon2003potassium}. Analogous to how calcium ions function in presynaptic terminals, these ions travel through ion channels to reach their targets and initiate biological processes\cite{rodriguez2016bioinspired}. Consequently, employing the molecular communication paradigm to model ion transport could yield novel insights into these critical mechanisms as well.

    \subsection{Vesicles}
    Vesicles, especially liposomes and extracellular vesicles (EV), are extensively investigated for drug delivery\cite{van2022liposomes, liang2021engineering,fonseca2021predatorprey}. Liposomes are synthetic lipid-shell containers that could be used to hold and deliver cargo such as drugs and proteins. Similarly. EVs, including microvesicles and exosomes, are natural lipid-shell nanoparticles that are derived from cells and are involved in many pathological processes such as cancer, infectious diseases, and neurodegenerative disorders\cite{el2013extracellular}. For the purpose of molecular communication, the vesicles could be used to contain drugs or other information carriers such as DNA and protein. Upon reception of the vesicles, the cargo will be released and information is transferred in the form of biochemical reactions. Furthermore, the transmission of the vesicles themselves in the body can be modeled using the molecular communication paradigm. For example in \cite{rudsari2022endtoend}, an exosome-based drug delivery channel is modeled, where the transmission phase of these exosomes operates through calcium-based exocytosis, mirroring the mechanism that vesicles employ for neurotransmitter release. The propagation is through free diffusion within the extracellular space, and the reception is based on biological processes such as ligand-receptor interactions or Clathrin-mediated endocytosis.
    

    \subsection{Proteins}
    Proteins contained in vesicles are commonly present in different signaling pathways. For instance, the signaling pathway between the Endoplasmic Reticulum (ER) and the Golgi apparatus, a shipping and regulating system for protein, is based on proteins contained in vesicles\cite{lee2004bi} and in \textit{Streptomyces coelicolor}, a bacterium, various proteins are found to be contained in membrane vesicles and are involved in biological processes such as metabolic processes and stress response\cite{faddetta2022streptomyces}. Moreover, in \cite{awan2019characterizing}, freely diffusing proteins are considered as the carriers of mechanosensitive signals, and a molecular communication channel model is established accordingly in the paper. Besides, as mentioned before, information can propagate on fluorescent proteins by FRET\cite{kuscu2011physical}, and the florescent proteins are commonly used as reporter molecules in receivers of molecular communication channels: upon the reception of the information molecules, the florescent protein is triggered and emits light to indicate the reception.

    \subsection{Phosphopeptides}
    Phosphopeptides, which are peptides containing one or more phosphorylated amino acids, are proposed as promising information molecules in molecular communication, as discussed in \cite{soldner2020survey}. These peptides undergo phosphorylation, a process where a phosphate group is added, occurring naturally within cells and also being inducible artificially for research purposes. There are several reasons why phosphopeptides are considered advantageous as information molecules. Firstly, phosphorylation is a common occurrence in cellular signaling, making phosphopeptides biocompatible. Additionally, compared to phosphorylated proteins, phosphopeptides are smaller and therefore diffuse faster, which is beneficial for molecular communication. Artificial synthesis of phosphopeptides is feasible, and notably, different phosphopeptides can interact with various proteins, providing an additional degree of freedom for type-based modulation in communication systems. However, it is important to note that, as of now, there are no extensive theoretical models or experimental testbeds developed further to investigate the use of these molecules in practical applications.

    \subsection{Sugar}
    Sugar, particularly glucose, is a focal point of research because of its prevalence in the human body and its crucial role in conditions such as diabetes. The glucose-insulin system is modeled as a molecular communication channel in \cite{abbasi2017information} to quantify and investigate the performance of the system from a communication theory perspective. The study in \cite{theodoridis2023glucose} proposes an in-body network capable of sensing glucose levels and releasing insulin from artificial beta cells, a critical function for regulating glucose levels. This work emphasizes the necessity of a detailed understanding of glucose transmission from a molecular communication standpoint. Further, \cite{koo2020deep} explores the development of a compact biosensor chip designed for implantation beneath the skin that continuously monitors glucose concentrations. Machine learning techniques are applied to enhance the accuracy and reliability of the sensor. Additionally, \cite{amerizadeh2021bacterial} details an experimental testbed using L-rhamnose, a specific type of sugar, as the information molecule. In this system, a bacterial receiver responds to L-rhamnose presence by quickly producing green fluorescent protein (GFP).
    
    \subsection{Polystyrene Microbeads}
    Polystyrene microbeads are spherical plastic particles typically several micrometers in size, renowned for their smoothness, uniformity, and resistance to degradation. These properties make them ideal for precise applications in scientific and engineering research. In \cite{duzyol2023microfluidic}, the researchers construct a microfluidic platform for molecular communication employing these polystyrene microbeads. The platform's distinctive feature allows for the tracking of each bead via sophisticated video processing techniques.

    \subsection{Cyclic Adenosine Monophosphates (cAMPs)}
    cAMP shares a role with calcium ions as a messenger in numerous biological processes. For instance, in the amoeba Dictyostelium discoideum (Dicty), cAMP is released under conditions of food scarcity to summon nearby Dicty cells for aggregation \cite{singer2019oscillatory}. The signal transduction involved in this response is modeled in \cite{hou2018signal}. Furthermore, \cite{wang2022biologically} proposes a rotational model of nanomachines, emulating the clustering behavior of Dicty. This innovative approach uses the rotational structure to enhance the system's robustness in noisy environments.

    \subsection{Acyl-homoserine Lactones (AHLs)}
    AHLs are large, complex molecules integral to \textit{quorum sensing}, a sophisticated mechanism permitting bacteria to communicate and synchronize behavior \cite{bassler1999how}. The study in \cite{einolghozati2013design} models a communication channel among engineered bacteria. This model operates on the principle that, upon receipt of AHLs, the bacteria respond by producing Green Fluorescent Protein (GFP), a luminescent marker facilitating communication with external devices. The channel relies on free diffusion, employing a concentration-based approach for modulation.

    \subsection{Isopropyl $\beta$-D-1-thiogalactopyranoside (IPTG)}
    IPTG is a structural mimic of lactose, which is widely utilized in molecular biology, serving as an inducer for gene expression within the lac operon of bacteria, analogous to the role of AHL in molecular communication. In the study cited as \cite{sezgen2021multiscale}, researchers construct an experimental platform wherein IPTG initiates the synthesis of red fluorescent protein in engineered bacteria, an activity monitored by a photodiode. This communication scheme is envisaged as an interface between in-body biosensors and external smart devices within Body Area Networks (BAN).

    \section{Future directions}
    With the field of molecular communication continuing to gain more momentum and the development of relevant fields such as synthetic biology and material science, the transition from theoretical studies to practical applications opens up many research opportunities. In this section, we propose several open research directions that are derived from the perspective of the information molecules.
        \subsection{Exploring More Information Molecules}
        In this paper, we introduce a range of information molecules. However, many, such as polystyrene microbeads, quantum dots, and phosphopeptides, remain underexplored. These molecules hold great potential as information carriers in nanonetworks, but there is a lack of theoretical modeling for their communication channels and experimental testbeds based on them. Similarly, while there are various types of neurotransmitters, ions, and odor molecules, only a few have been investigated within the molecular communication paradigm. These molecules perform crucial functions in different parts of various organisms. Studying them from an ICT perspective could provide significant insights into biological mechanisms, leading to applications in diverse fields like medicine and agriculture.
        
        \subsection{Incorporating Physical Characteristics in Channel Models}
        Although various channel models have been developed for different information molecules, many of their physical characteristics have not yet been fully explored. For instance, the size of manufactured Superparamagnetic Iron Oxide Nanoparticles (SPIONs) is not uniformly consistent but typically follows a distribution\cite{kiss1999new}. This variance can introduce additional noise in the communication channel. Similarly, the topology of DNA molecules affects their diffusivity, which in turn alters the communication channel model \cite{robertson2006diffusion}. Numerous other physical characteristics, as discussed, can be integrated to develop more realistic channel models. This becomes increasingly important as experimental testbeds become more prevalent and are employed in more intricate scenarios.
        
        \subsection{Comparing with Existing Data Sets}
        Although numerous theoretical channel models have been developed, only a few have been compared with existing biological datasets. As mentioned previously, one of the primary purposes of modeling biological communication is to gain a deeper understanding of the underlying mechanisms. Furthermore, these models can be applied to pathological conditions for the diagnosis and treatment of diseases. Specifically, this approach is particularly useful for synaptic communication and neurological diseases, given the relative abundance of data on neuronal activity. Additionally, this method could be applied to the study of information molecules involved in signaling pathways, such as ions and proteins.
        
        \subsection{Implementing the Information Molecules}
        Implementing information molecules in practical applications requires efforts across multiple dimensions. For the design of the transmitter, as illustrated in Figure \ref{fig: transmitter}, a source for information molecules, a processing unit, and a release mechanism need to be developed. The functions of the receivers also require further consideration. For example, they could serve as interfaces to external devices, relay points for further transmission, or initiators for responses such as drug release or directional changes in targeted drug delivery. These all would depend on the specific relevant information molecules. Additionally, the miniaturization, energy harvesting, and biocompatibility of these devices are critical factors that need thorough investigation as well.
        
        \subsection{Specifying Application Scenarios}
        In the existing literature on molecular communication, although a broad discussion of application scenarios exists, these need more detailed investigation. In \cite{chude2017molecular}, a classification of scenarios is proposed, encompassing cardiovascular, extracellular space/cell surface, intracellular, whole-body, and nervous signaling channels. This classification extends to macroscale communication in environments like mines, underwater, and agriculture, demonstrating that molecular communication applications span a wide range of fields. Therefore, firstly, more specific classifications could be developed for these application scenarios. For instance, the molecular communication channels in the brain, cardiovascular system, and other organs would have distinct channel properties due the their biological differences. Moreover, while the suitability of information molecules for these various scenarios is briefly discussed in this paper, more substantial research is also needed.
        
        \subsection{Designing Modulation Methods}
        The physical characteristics and channel properties of information molecules can be utilized to design modulation methods that are more robust and efficient. For instance, the shell of SPIONs and the topology of proteins and DNA can be employed in type-based modulation. This approach heavily depends on the design of the transmitters and receivers, which must provide sufficient distinguishing functionality.

        \subsection{Integration for Internet of Bio-Nano Things}
        Finally, with advancements in application scenarios, modulation techniques, and implementation strategies, it is conceivable that the Internet of Bio-Nano Things could be realized through molecular communication. This can be achieved by integrating relay and response micro/nanorobots within the body, utilizing various information molecules tailored to different body parts. Furthermore, interfaces capable of converting in-body molecular signals into electrical signals for external devices can be developed. This conversion can use specific detection methods of information molecules, such as exploiting the magnetism of SPIONs or utilizing Green Fluorescent Protein (GFP) for light signal transmission.

    \section{Conclusion}
    In this paper, we have provided a comprehensive overview of the information molecules featured in existing molecular communication literature, detailing their physical characteristics, communication channel properties, communication performance, and application scenarios. Given the generality of molecular communication, this paradigm encompasses a broad range of molecules. As demonstrated in our discussion, information molecules exhibit significant diversity in the aforementioned properties, necessitating individualized consideration for the development of realistic models and practical applications. Moreover, certain properties of information molecules remain underexplored, presenting potential advantages or drawbacks for their respective applications.

    We propose that the significance of the information molecules themselves should not be understated. To facilitate the transition from theoretical research to practical application, it is imperative to more thoroughly account for these  properties of the molecules. By offering a comprehensive overview of information molecules, this survey aims to equip researchers in the field with fundamental information of them and spur further studies that integrate their characteristics into theoretical models, experimental testbeds, and ultimately, real-world applications.


\begin{thebibliography}{100}
\bibitem{figure}
This figure and the following biological illustrations are created with
  BioRender.com.
\end{thebibliography}
\end{document}